\documentclass[12pt,draftclsnofoot,onecolumn,journal,letterpaper]{IEEEtran}
\usepackage{amssymb}
\usepackage{algorithm}
\usepackage{polynomial}

\usepackage{color}
\usepackage[dvipsnames]{xcolor}
\usepackage{algpseudocode}
\usepackage{cite}
\usepackage{graphicx}
\usepackage{esdiff}
\usepackage{amsthm}
\usepackage{relsize}

\theoremstyle{plain}
\newtheorem{theorem}{Theorem}

\theoremstyle{definition}

\usepackage{amsfonts}
\usepackage[cmex10]{amsmath}
\ifCLASSOPTIONcompsoc
\usepackage[tight,normalsize,sf,SF]{subfigure}
\else
\usepackage[tight,footnotesize]{subfigure}
\fi

\usepackage{amsmath}
\usepackage{epstopdf}
\usepackage{bbm}
\usepackage{array}
\usepackage{mdwmath}
\usepackage{mdwtab}
\usepackage{multirow}
\usepackage{hyperref}
\usepackage{mathtools}
\usepackage{nicefrac}
\makeatletter
\newcases{mycases}{\quad}{%
  \hfil$\m@th\displaystyle{##}$}{$\m@th\displaystyle{##}$\hfil}{\lbrace}{.}
\makeatother
\renewcommand{\algorithmicrequire}{\textbf{Input:}}

\begin{document}

\title{Rate Adaptation and Admission Control for Video Transmission with Subjective Quality Constraints}
\author{Chao Chen,~\IEEEmembership{Student Member,~IEEE,}
        Xiaoqing Zhu,~\IEEEmembership{Member,~IEEE,}
        Gustavo~de~Veciana,~\IEEEmembership{Fellow,~IEEE,}
        Alan C. Bovik,~\IEEEmembership{Fellow,~IEEE,}\\
        Robert W. Heath Jr.,~\IEEEmembership{Fellow,~IEEE}

\thanks{Chao Chen, Gustavo~de~Veciana, Alan Bovik, and Robert Heath are with Department of Electrical and Computer Engineering,
The University of Texas at Austin, Austin, TX 78712-0240, USA (email: tochenchao@gmail.com; gustavo@ece.utexas.edu; bovik@ece.utexas.edu; rheath@ece.utexas.edu). Xiaoqing Zhu is with the Advanced Architecture and Research Group, Cisco Systems, San Jose, CA 95134, USA (email: xiaoqzhu@cisco.com).

This research was supported in part by Intel Inc. and Cisco Corp. under the VAWN program.}
}

\maketitle

\begin{abstract}
Adapting video data rate during streaming can effectively reduce the risk of playback interruptions caused by channel throughput fluctuations. The variations in rate, however, also introduce video quality fluctuations and thus potentially affects viewers' Quality of Experience (QoE). We show how the QoE of video users can be improved by rate adaptation and admission control. We conducted a subjective study wherein we found that viewers' QoE was strongly correlated with the empirical cumulative distribution function (eCDF) of the predicted video quality. Based on this observation, we propose a rate-adaptation algorithm that can incorporate QoE constraints on the empirical cumulative quality distribution per user. We then propose a threshold-based admission control policy to block users whose empirical cumulative quality distribution is not likely to satisfy their QoE constraint. We further devise an online adaptation algorithm to automatically optimize the threshold. Extensive simulation results show that the proposed scheme can reduce network resource consumption by 40\% over conventional average-quality maximized rate-adaptation algorithms.
\end{abstract}


%
\IEEEpeerreviewmaketitle
\section{Introduction}
\label{sec:intro}
\IEEEPARstart{V}{ideo} traffic is currently a rapidly growing fraction of the data traffic on wireless networks. As reported in \cite{CiscoVNI}, video traffic accounted for more than 50\% of the mobile data traffic in 2012. Efficiently utilizing network resources to satisfy video users' expectations regarding their Quality of Experience (QoE) is an important research topic. In this paper, we study approaches to share wireless down-link resources among video users via QoE-based rate adaptation and admission control.

We focus on a setting in which stored video content is streamed over wireless networks. When a video is streamed, the received video data is first buffered at the receiver and then played out to the viewer. Because the throughput of a wireless channel generally varies over time, the amount of buffered video decreases when the channel throughput falls below the current video data rate. Once all the video data buffered at the receiver has been played out, the playback process stalls, significantly affecting the viewers QoE \cite{ZhangHui_QoE}. To address this problem, various video rate-adaptation techniques based on scalable video coding or adaptive bitrate switching have been proposed to match the video data rate to the varying channel capacity \cite{SVC,smoothstream,livestream,dynamicsstream,DASH}. Although these rate adaptation techniques can effectively reduce the risk of playback interruptions, the variable bitrate causes quality fluctuations, which, in turn, affect viewers' QoE. In most existing rate-adaptation algorithms such as \cite{LunaJSAC03,ChaTMM06,HuangCSVT08,xiaoqingCSVT10,LinTMC13,HaoTMM13}, the average video quality is employed as the proxy for QoE. The average quality, however, does not reflect the impact of quality fluctuations on the QoE, i.e., two videos with the same average quality can have significantly different levels of quality fluctuation. In this paper, we propose to characterize and predict the users' QoE using the second order empirical cumulative distribution function ($2^\mathrm{nd}$-order eCDF) of the delivered video quality; this is defined as
\begin{align}
\label{eq:f2}
\mathrm{F}^{(2)}(x;\mathrm{q})=\frac{1}{T}\sum_{t=1}^T\max\left\{x-\mathrm{q}(t),0\right\},
\end{align}
where $\mathrm{q}(t)$ represents the predicted quality \cite{AutomaticVQA} of the $t^\mathrm{th}$ second of the video and $T$ is the video length. Note that $\max\{x-\mathrm{q}(t),0\}>0$ if and only if $\mathrm{q}(t)< x$, so the $2^\mathrm{nd}$-order eCDF captures for how long and by how much the predicted video quality falls below $x$. If we interpret $x$ as the threshold below which users judge the video quality to be unacceptable, then the $2^\mathrm{nd}$-order eCDF reflects the impact of the unacceptable periods on the QoE. Since it has been recognized that the worst parts of a video tend to dominate the overall quality of an entire video \cite{Barkowsky07,Ninassi09,NQA10,SesBov11,VQPooling}, the $2^\mathrm{nd}$-order eCDF can be used to predict the QoE.

The efficacy of the $2^\mathrm{nd}$-order eCDF in capturing QoE can be validated through subjective experiments. In \cite{chaoICASSP} and \cite{report_dynamic_model}, we reported a subjective study of this type. For each of the 15 quality-varying videos involved in the subjective study, we asked 25 subjects to score its quality. We computed the linear correlation coefficients (LCCs) between various QoE metrics and the subjects' Mean Opinion Scores (MOSs) in Table~\ref{tab:correlation table}.
The $2^\mathrm{nd}$-order eCDF achieves the strongest linear correlation (0.84). In comparison, the average video quality only achieves a correlation of 0.57. This lends strong support for eCDF as a good proxy for video QoE.

\begin{table}[!h]
\caption{The Linear Correlation Coefficients (LCCs) of Several Metrics with QoE. The metrics $\mathrm{mean}\{\mathrm{q}\}$, $\mathrm{min}\{\mathrm{q}\}$ and $\mathrm{var}\{\mathrm{q}\}$ are the average value, the maximum value and the variance of the time series $\mathrm{q}(1),\dots,\mathrm{q}(T)$, where $\mathrm{q}(t)$ is the predicted video quality at time $t$.}
\label{tab:correlation table}
\centering
\begin{tabular}{c|c|c|c|c}
\hline
\bfseries QoE metrics & $\mathrm{mean}\{\mathrm{q}\}$ & $\min\{\mathrm{q}\}$&$\mathrm{mean}\{\mathrm{q}\}+\mu\sqrt{\mathrm{var}\{\mathrm{q}\}}$&$\mathrm{F}^{(2)}(x;\mathrm{q})$ \\
\hline
\bfseries LCC&0.5659&0.4022&0.7377&0.8446\\
\hline
\end{tabular}
\end{table}
In this paper, we design adaptive video streaming algorithms that incorporate QoE constraints on the $2^\mathrm{nd}$-order eCDFs of the video qualities seen by users. In particular, we consider a wireless network in which a base station transmits videos to multiple users. The user population is dynamic, i.e., users arrive and depart from the network at random times. When a new user joins the network, the base station starts streaming a video to the user. A rate adaptation algorithm is employed to control the video data rate of all active video streams according to varying wireless channel conditions.

When the base station is shared by too many users simultaneously, the QoE served to each user can be poor. Instead of serving every user with poor QoE, it is preferable to satisfy the QoE constraints of existing users by selectively blocking newcomers. Therefore, in addition to rate adaptation algorithms, we introduce a new admission control strategy that is designed to maximize the number of video users satisfying the QoE constraints on their $2^\mathrm{nd}$-order eCDFs. As will be shown in the paper, although the admission control strategy damages the QoE of the blocked users, the overall percentage of users satisfying the QoE constraints among both admitted and blocked users can be significantly improved. The contributions of our work are twofold:

\begin{enumerate}
\item{\it An online rate-adaptation algorithm aimed at meeting QoE constraints based on the $2^\mathrm{nd}$-order eCDFs of users' video qualities.} \\Since the $2^\mathrm{nd}$-order eCDF is determined by the overall spatio-temporal pattern of video, simply maximizing the video quality all the times is not sufficient to satisfy the QoE constraints. Instead, we propose an online rate-adaptation algorithm that jointly considers the channel conditions, the video rate-quality characteristics, and the $2^\mathrm{nd}$-order eCDFs of all video users. We show significant performance gains over conventional average-quality optimized algorithms.
\item{\it An admission control strategy, which blocks video users who will likely be unable to meet the constraints on their $2^\mathrm{nd}$-order eCDFs.} \\Specifically, we propose an algorithm that predicts the video quality experienced by each user who is new to the network. We then employ a threshold-based admission control policy to block those users whose estimated qualities fall below the threshold. An online algorithm is proposed to adjust the threshold to its optimal value. The proposed admission control strategy further improves the performance of our rate-adaptation algorithm, especially when the network resources are limited.
\end{enumerate}
The remainder of this paper is organized as follows: Section~\ref{sec:relatedwork} discusses related work. In Section~\ref{sec:overview}, we give an overview of the structure of the video streaming system studied in this paper. In Section~\ref{sec:sysmodel}, we explain our system model and the QoE constraints. The proposed rate adaptation algorithm and the admission control strategy are introduced in Section~\ref{sec:caseI} and \ref{sec:case2}. We demonstrate their performance via extensive simulation results. Section \ref{sec:conclusion} concludes the paper with discussions on future work.

\section{Related Work}
\label{sec:relatedwork}
Let us first review related work in QoE assessment, rate adaptation, and admission control.\\
\subsection{QoE Metrics for Rate-Adaptive Video Steaming:}
Most existing rate-adaptation algorithms such as \cite{LunaJSAC03,ChaTMM06,HuangCSVT08,xiaoqingCSVT10,LinTMC13,HaoTMM13} employ average video quality as the proxy of QoE due to its simplicity in analysis. As shown in Table~\ref{tab:correlation table}, however, average quality does not efficiently capture the impact of quality fluctuations. To address this problem, temporal quality pooling has been studied \cite{Barkowsky07,Ninassi09,NQA10,SesBov11,VQPooling}. The pooling algorithms, however, have only been designed and validated for short videos that are a few seconds long. Furthermore, they are too complicated to be incorporated into a tractable analytical framework for the design of rate-adaptation algorithms. The authors of \cite{Yim2011} propose a simple QoE metric, which is the weighted sum of the time-average and the standard deviation of the predicted video quality. As shown in Table~\ref{tab:correlation table}, this metric achieves a better correlation of 0.73 than the time-average quality. The QoE metric that we propose here achieves an even higher correlation of 0.84. Moreover, because the $2^\mathrm{nd}$-order eCDF is simply a temporal average of the function $\max\{x-\mathrm{q}(t),0\}$, its analysis is just as simple as for average video quality.


\subsection{Rate Adaptive Video Streaming:}
Extensive research efforts have been applied to the problem of rate adaptation for wireless video streaming. Most existing work employs the time-average video quality as the QoE metric due to its simplicity \cite{LunaJSAC03,ChaTMM06,HuangCSVT08,xiaoqingCSVT10,LinTMC13,HaoTMM13}. In \cite{VinayInfocom12}, a rate adaptation algorithm is proposed to optimize the QoE metric of \cite{Yim2011}. Although the proposed algorithm mitigates video quality variations, it assumes a fixed set of users and thus does not incorporate the impact of user arrival and departure on video quality. In \cite{WeberTON05}, the quality fluctuations caused by user arrival and departure are analyzed for large wireless links shared by many video streams. The algorithms proposed here do not rely on such assumptions and can be applied to small wireless cells such as Wi-Fi networks.

\subsection{Admission Control for Video Streaming:}
Admission control for variable bitrate videos has been studied in \cite{CAC98,CAC05,CAC08TMM}. In \cite{CAC98}, an admission control algorithm is proposed for variable bitrate videos stored on disk arrays and transmitted over cable television networks. In contrast to wireless networks, the transmission bottleneck of cable television networks is the buffer at the disk array. The admission control algorithm is designed to ensure that the buffer does not overflow while the video is played back continuously. In \cite{CAC05} and \cite{CAC08TMM}, threshold-based admission control algorithms for video streams delivered over packet-switch networks are studied. In \cite{CAC05}, thresholds are applied to the number of video users. The heterogeneity in the data rate of the video streams is not considered. In \cite{CAC08TMM}, the threshold is applied to the aggregated data rate requested by admitted videos. In both \cite{CAC05} and \cite{CAC08TMM}, a statistical model of the video traffic is necessary to optimize the admission threshold. In practice, however, modeling the video traffic is difficult. Our proposed admission control strategy does not rely on the \emph{a priori} knowledge of the traffic statistics. The threshold is learned online and optimized automatically.

\section{System Overview}
\label{sec:overview}
We first discuss the architecture of the wireless networks considered in this paper. Then, we explain how the proposed online rate adaptation algorithm and threshold-based admission control strategy fit into the existing network architecture.
\subsection{Architecture of the Wireless Network}
\begin{figure}[h!]
\centering
\includegraphics[width=0.8\columnwidth]{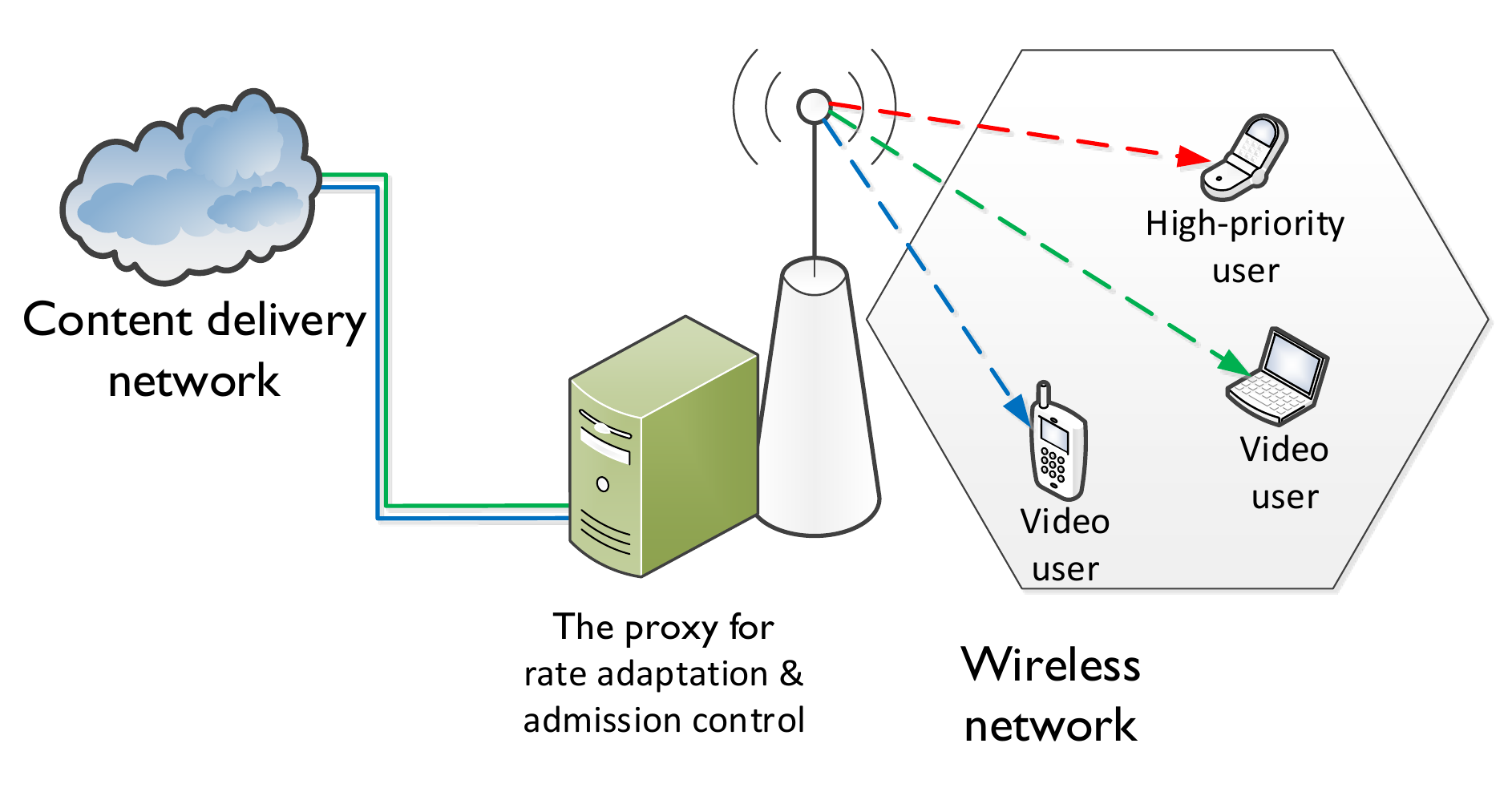}
\caption{The wireless network considered here: The video is stored at the content delivery network. The proxy for rate adaptation and admission control is colocated with the base station. The base station serves both video users and high-priority users.}
\label{fig:sysmodel}
\end{figure}
We consider a wireless network where video users share the down-link with high-priority traffic (e.g., voice traffic) and thus video users will need to adapt their rates accordingly.
(see Fig.~\ref{fig:sysmodel}). All users arrive to and depart from the network at random times. A high-priority user requires a random amount of wireless resource (e.g., transmission time in TDMA systems) per unit time throughout its sojourn in the system. The transmission resources not used by the high-priority  users can be allocated to video users. When a video user arrives, it requests a video that is stored at a content delivery network (CDN) and is streamed to the user via the base station. When a video is being streamed, the video data is first delivered to a receive buffer and then decoded for display. Paralleling prior work such as \cite{HaoTMM13}\cite{VinayInfocom12}\cite{PuPVW12}, we assume a proxy is colocated with the base station. We treat the high-priority users as background traffic, and the proxy is only used to control the video streams.

\subsection{Video Streaming Proxy}

\begin{figure}[h!]
\centering
\includegraphics[width=0.8\columnwidth]{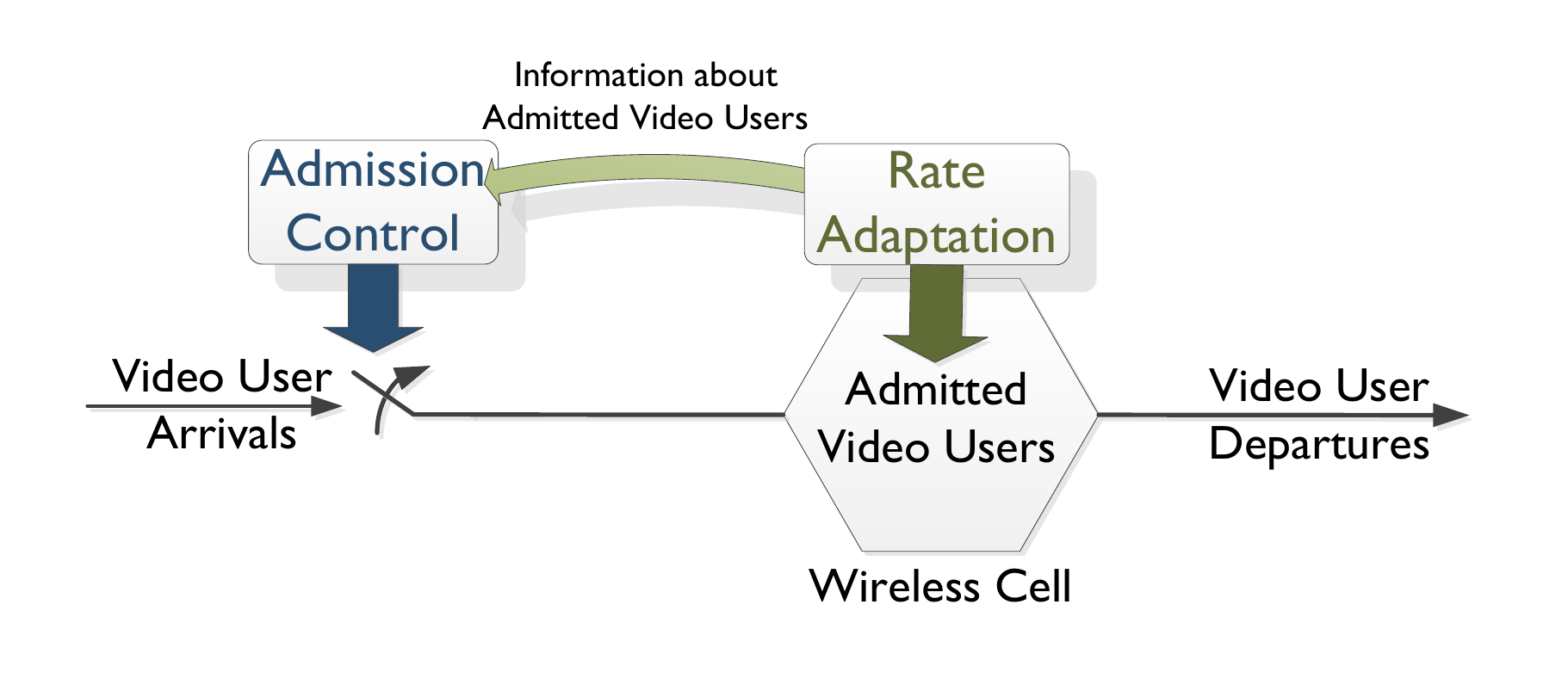}
\caption{The proposed QoE-constrained video streaming system: Admission control only affects newly arrived video users and the rate adaptation algorithm does its best to satisfy the QoE constraints of all admitted video users.}
\label{fig:block1}
\end{figure}

The function of the proxy is twofold (see Fig.~\ref{fig:block1}). First, when a video user arrives, the proxy decides whether the user should be admitted to share the channel or not. Second, for admitted video users, the proxy adapts the transmission data rates according to the varying channel conditions. The admission control strategy only acts on newly arrived video users, and the rate adaptation algorithm does its best to satisfy QoE constraints for all admitted video users. The rate adaptation algorithm accompanies the admission control algorithm by feeding back necessary information (see Section \ref{sec:admisson_case1} for more detail) regarding the current status of the admitted users.

We assume the proxy operates in a time-slotted manner where the duration of each slot is $\Delta T$ seconds. The admission control and the rate adaptation actions are conducted at the beginning of every time slot. Note that, in a video stream, the video frames are partitioned in Groups of Pictures (GoPs) and the video data rate can only be adapted at the boundary of GoPs \cite{H264}\cite{H265}. Because the duration of a GoP is usually configured to be larger than one second to achieve high compression efficiency, we assume $\Delta T$ is at least 1 second.

\section{System Model}
\label{sec:sysmodel}
Before proceeding further, we introduce the notation used in the paper. Then, we describe the channel model and the video rate-quality model. At the end of this section, we explain the QoE constraints considered in our problem formulation.
\subsection{Notation}
In the rest of the paper, the time slots are indexed with $t=1,2,\dots$. The notation $\left(\mathrm{x}(t),~t=1,2\dots\right)$ denote a discrete time series. Lower-case symbols such as $a$ denote scalar variables and boldface symbols such as $\mathbf a$ denote vectors. Random variables are denoted by uppercase letters such as $\mathsf{A}$. Calligraphic symbols such as $\mathcal{A}$ denote sets, while $|\mathcal A|$ is the cardinality of $\mathcal{A}$. The set of positive integers is denoted  $\mathbb{N}^+$. The set of real numbers is denoted $\mathbb{R}$. Finally, the function $[x]^+=\max\{x,0\}$.
\subsection{Wireless Channel Model}
\label{sec:channelmodel}
We label users (including both video users and high-priority users) according to their arrival times, i.e., user $u$ is the $u^\mathrm{th}$ user to arrive to the network. For each user $u$, we let $\mathsf{A}_u$ and $\mathsf{D}_u$ be random variables denoting the arrival and departure times, respectively. The time spent by a user in the network is denoted by $\mathsf{T}_u=\mathsf{D}_u-\mathsf{A}_u+1$. 

We let by $\mathcal{U}^\mathrm{p}(t)$ be the set of high-priority users in the network at slot $t$. For each high-priority user $u\in\mathcal{U}^\mathrm{p}(t)$, we let the random variable $\mathsf{W}_u(t)$ represent the amount of data received in slot $t$. The data rate is thus $\mathsf{R}_u(t)=\mathsf{W}_u(t)/\Delta T$. We denote by $\mathcal{U}^\mathrm{v}(t)$ the set of video users that would be in the network at slot $t$ if all were admitted. The set of video users in $\mathcal{U}^\mathrm{v}(t)$ who are actually admitted to share the wireless channel is denoted by $\mathcal{U}^\mathrm{av}(t)$. We assume that an admission decision is made upon the arrival of each video user. Once admitted, the video user shares the channel until it leaves the network. For an admitted video user $u\in\mathcal{U}^\mathrm{av}(t)$, we denote by $\mathrm{w}_u(t)$ the amount of received video data in slot $t$. The video data rate delivered to the user is thus $\mathrm{r}_u(t)=\mathrm{w}_u(t)/\Delta T$. We call $\mathbf{r}(t)=\left(\mathrm{r}_u(t):~u\in\mathcal{U}^\mathrm{av}(t)\right)$ the {\it video rate vector} at time $t$. Because high-priority users are treated as background traffic, the proxy only controls the video rate vector and regards $\left(\mathsf{R}_u(t):~u\in\mathcal{U}^\mathrm{p}(t)\right)$ as exogenous variables.


We assume the set of video rate vectors that can be supported by the wireless channel is given by
\begin{align}
\label{eq:channelmodel}
\mathcal{C}(t)=\{\mathbf{r}: \mathrm{C}_t(\mathbf{r})\leq 0\},
\end{align}
where
$\mathrm{C}_t : \mathbb{R}^{|\mathcal{U}^\mathrm{av}(t)|}\rightarrow \mathbb{R}$ is a time-varying multivariate convex function. The specific form of $\mathrm{C}_t(\cdot)$ depends on the multiuser multiplexing techniques used in the wireless network. For example, in a time-division multiple access (TDMA) system \cite{HaoTMM13}\cite{VinayInfocom12}, the channel is occupied by a single user at any moment. Denote by $\mathsf{P}_u(t)$ the peak transmission rate of user $u$, i.e., the transmission rate at which user $u$ can be served during slot $t$. Then, in slot $t$, video user $u\in\mathcal{U}^\mathrm{av}(t)$ spends $\frac{\mathrm{w}_u(t)}{\mathsf{P}_u(t)}$ seconds to download video. Similarly, each high-priority user $u'\in\mathcal{U}^\mathrm{p}(t)$ expends $\frac{\mathsf{W}_{u'}(t)}{\mathsf{P}_{u'}(t)}$ seconds downloading data. Since the total transmission across all users is less than $\Delta T$, we have $\sum_{u\in\mathcal{U}^\mathrm{av}(t)}\frac{\mathrm{w}_u(t)}{\mathsf{P}_u(t)}+\sum_{{u'}\in\mathcal{U}^\mathrm{p}(t)}\frac{\mathsf{W}_{u'}(t)}{\mathsf{P}_{u'}(t)}\leq\Delta T$. Because $\mathrm{r}_u(t)=\mathrm{w}_u(t)/\Delta T$ and $\mathsf{R}_{u'}(t)=\mathsf{W}_{u'}(t)/\Delta T$, we have $\sum_{u\in\mathcal{U}^\mathrm{av}(t)}\frac{\mathrm{r}_u(t)}{\mathsf{P}_u(t)}+\sum_{{u'}\in\mathcal{U}^\mathrm{p}(t)}\frac{\mathsf{R}_{u'}(t)}{\mathsf{P}_{u'}(t)}\leq 1$. Therefore, for TDMA systems, we have $\mathrm{C}_t\left(\mathbf{r}(t)\right)=\sum_{u\in\mathcal{U}^\mathrm{av}(t)}\frac{\mathrm{r}_u(t)}{\mathsf{P}_u(t)}+\sum_{{u'}\in\mathcal{U}^\mathrm{p}(t)}\frac{\mathsf{R}_{u'}(t)}{\mathsf{P}_{u'}(t)}-1$.

In rate-adaptive video streaming systems such as \cite{DASH,dynamicsstream,livestream}, the video data rate can only take values in a finite and discrete set. In our problem formulation, we relax the constraint and allow $\mathrm{r}_u(t)$ to take values in a compact interval $[\mathrm{r}^\mathrm{min}_u(t),\mathrm{r}^\mathrm{max}_u(t)]$, where $\mathrm{r}^\mathrm{min}_u(t)$ and $\mathrm{r}^\mathrm{max}_u(t)$ denote the minimum and maximum data rate available for user $u$, respectively. Letting $\mathcal{R}(t)=\Pi_{u\in\mathcal{U}^\mathrm{av}(t)}[\mathrm{r}^\mathrm{min}_u(t),\mathrm{r}^\mathrm{max}_u(t)]$, we have
\begin{align}
\label{eq:min_max}
\mathbf{r}(t)\in\mathcal{R}(t).
\end{align}
In our algorithm implementation, we round up the optimal video data rate obtained under this relaxed constraint to the nearest available data rate.


\subsection{Video Rate-Quality Model}
We assume that the quality of the video downloaded in each slot is represented by a Difference Mean Opinion Score (DMOS) \cite{ITU}, which ranges from 0 to 100 where lower values indicate better quality. To represent video quality more naturally, so that higher numbers indicate better video quality, we deploy a Reversed DMOS (RDMOS). Denote by $\mathrm{q}^\mathrm{dmos}_u(t)$ the DMOS of the video delivered to user $u$ at slot $t$, the RDMOS is given by $\mathrm{q}^\mathrm{rdmos}_u(t) =  100-\mathrm{q}^\mathrm{dmos}_u(t)$. We employ the following rate-quality model to predict $\mathrm{q}^\mathrm{rdmos}_u(t)$ using the video data rate $\mathrm{r}_u(t)$:
\begin{align}
\label{eq:ratequality}
\mathrm{q}_u(t)=\alpha_u(t)\log(\mathrm{r}_u(t))+\beta_u(t),
\end{align}
where $\mathrm{q}_u(t)$ is the predicted RDMOS. The model parameters $\alpha_u(t)$ and $\beta_u(t)$ can be determined by minimizing the prediction error between $\mathrm{q}^\mathrm{rdmos}_u(t)$ and $\mathrm{q}_u(t)$. For stored video streaming, the video file is stored at the CDN. Thus, the model parameters in \eqref{eq:ratequality} can be obtained before transmission. Here, we assume the parameters $\alpha_u(t)$ and $\beta_u(t)$ are known \emph{a prior}.

\begin{figure}[!h]
\centering
\includegraphics[width=0.6\columnwidth]{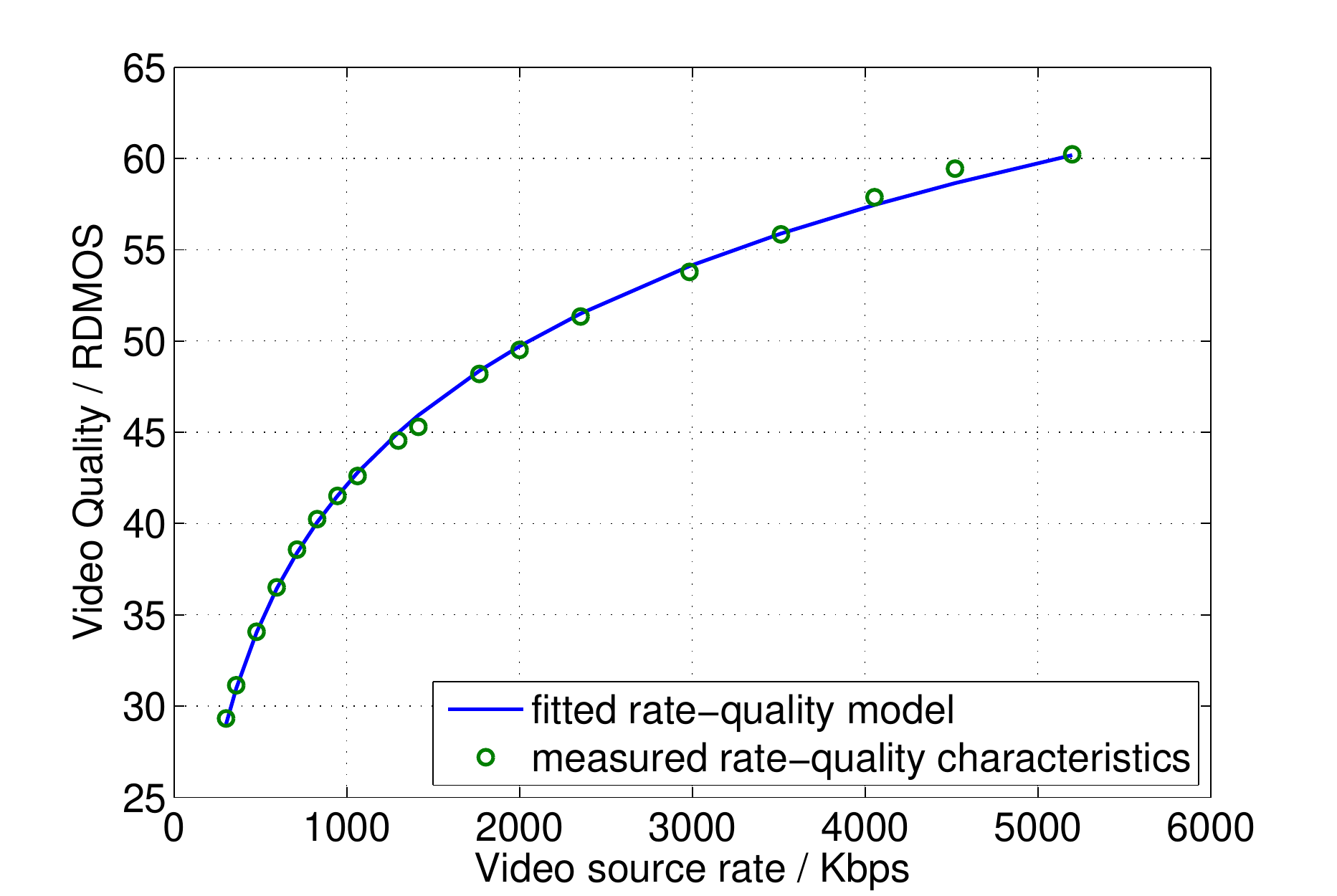}
\caption{The performance of the rate-quality model on one second of a video randomly chosen from the database \cite{chaoICASSP}. The rate-quality characteristics are shown in circles while the fitted rate-quality model \eqref{eq:ratequality} is the solid line.}
\label{fig:rd_log_model}
\end{figure}

We validated this model on a video database that includes twenty-five different pristine and representative videos \cite{chaoICASSP}. The rest of the database is created by encoding and decoding each video at different rates with the widely used H.264 codec \cite{ffmpeg}. Then, we predict the RDMOSs for each second of the decoded videos using the high-accuracy video-RRED index \cite{RRED}. In Fig.~\ref{fig:rd_log_model}, we show the rate-RDMOS mapping of one second of a video randomly chosen from the database. It may be observed that the model \eqref{eq:ratequality} can accurately predict the RDMOSs. On the whole database, the mean prediction error of \eqref{eq:ratequality} is less than 1.5, which is visually negligible. Thus, in the following, we shall refer to $\mathrm{q}_u(t)$ as the video quality.

\subsection{Constraints on the Quality of Experience}
\label{sec:constraints}

\begin{figure}[h!]
    \centering
    \subfigure[]{\includegraphics[width=.45\columnwidth]{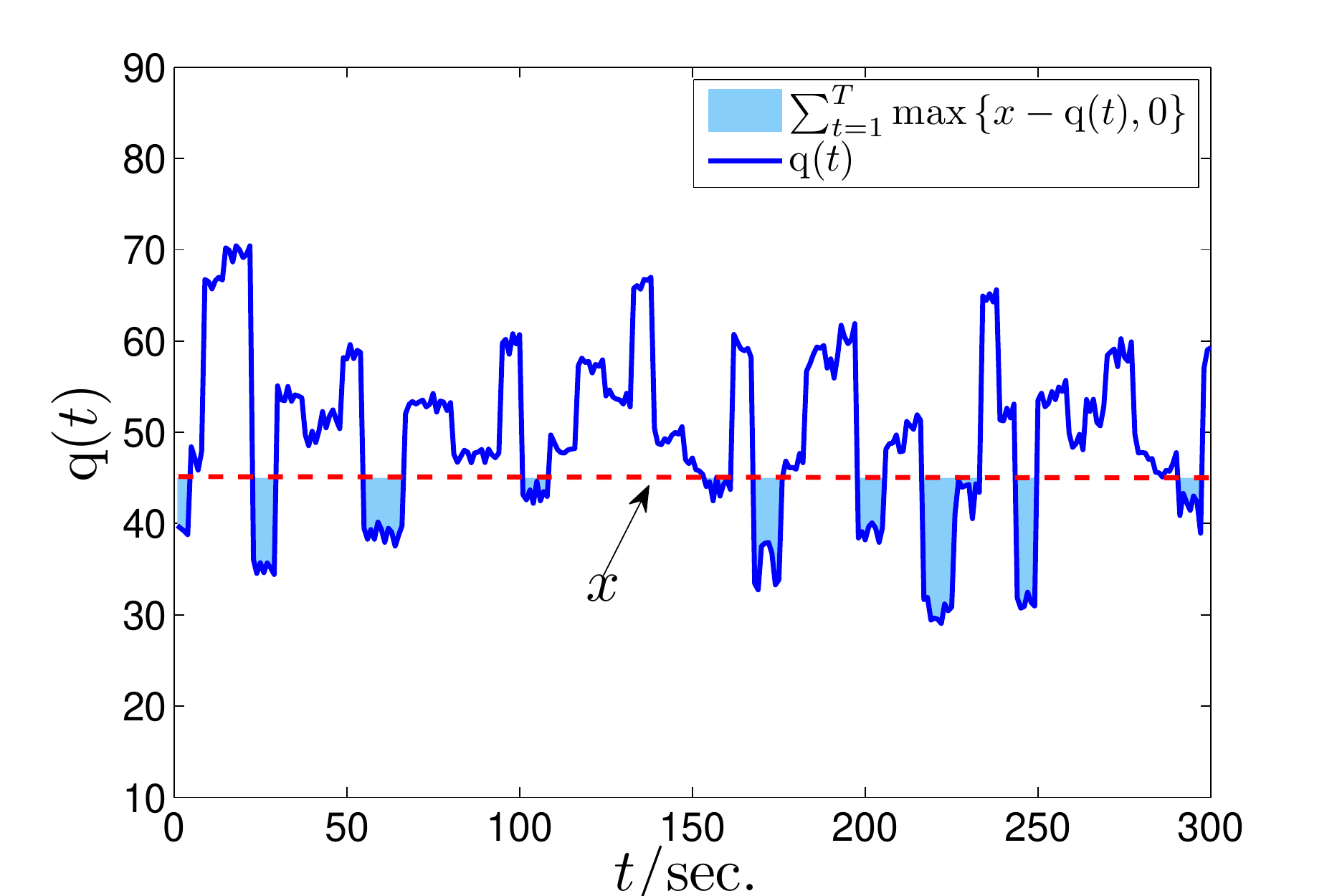}
    \label{fig:x_f2}}
    \subfigure[]{\includegraphics[width=.45\columnwidth]{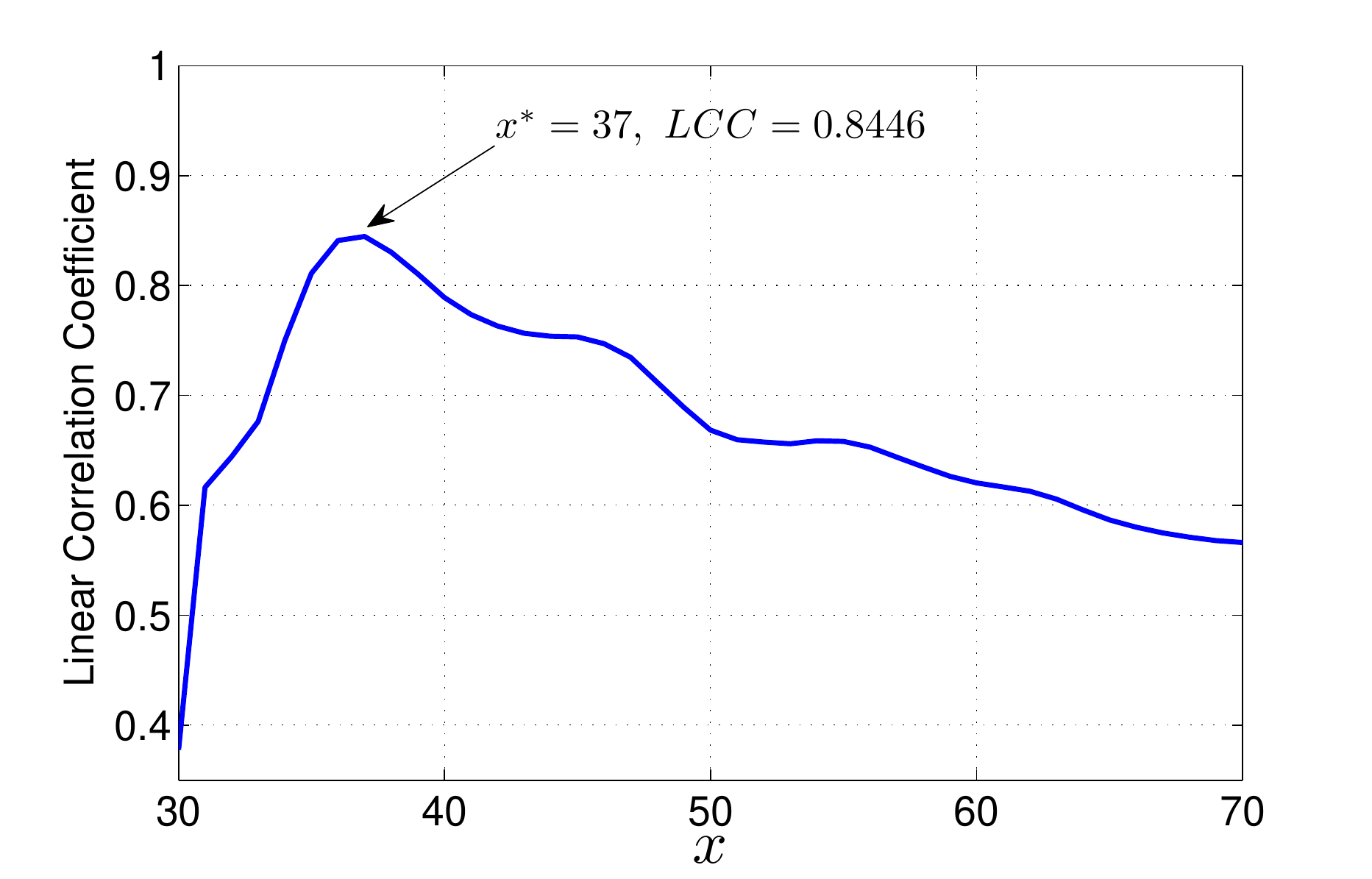}
    \label{fig:f2_lcc}}
    \caption{ \subref{fig:x_f2} An example of $\mathrm{F}^{(2)}(x;\mathrm{q})$ at $x=45$; \subref{fig:f2_lcc} The absolute value of the linear correlation coefficient (LCC) between $\mathrm{F}^{(2)}(x;\mathrm{q})$ and the mean opinion scores.}
    \label{fig:subjectivestudy}
\end{figure}

We capture video users' QoE using the $2^\mathrm{th}$-order eCDF $\mathrm{F}^{(2)}(x;\mathrm{q})$, which was defined in \eqref{eq:f2}. As illustrated in Fig.~\ref{fig:subjectivestudy}\subref{fig:x_f2}, for a given $x$, the right-hand side of \eqref{eq:f2} is proportional to the area where $\mathrm{q}(t)$ falls below $x$. If $\mathrm{q}(t)$ falls below $x$ for a long while, the QoE is poor and $\mathrm{F}^{(2)}(x;\mathrm{q})$ is large. Otherwise, the QoE is good and $\mathrm{F}^{(2)}(x;\mathrm{q})$ is small.

To justify the use of the $\mathrm{F}^{(2)}(x;\mathrm{q})$ as the QoE metric, we conducted a subjective study following the guidelines of \cite{ITU}. The study involved twenty-five subjects and fifteen quality-varying long videos (for more details, see \cite{chaoICASSP} and \cite{report_dynamic_model}). Based on the subjects' feedback, we obtained   the Mean Opinion Scores (MOSs) of each video's overall quality. Given an $x$, we computed  $\mathrm{F}^{(2)}(x;\mathrm{q})$ for all the videos in the database and then calculated the linear correlation coefficient (LCC) between the computed $\mathrm{F}^{(2)}(x;\mathrm{q})$s and the MOSs. In Fig~\ref{fig:subjectivestudy}\subref{fig:f2_lcc}, we plot the absolute value of the LCCs as a function of $x$. We found that, at $x^*=37$, $\mathrm{F}^{(2)}(x^*;\mathrm{q})$ achieves a strong correlation of 0.84 with the MOSs. Since $\mathrm{F}^{(2)}(x^*;\mathrm{q})$ is determined by the area where $\mathrm{q}(t)$ falls below $x^*$, we interpret $x^*$ as the users' {\it video quality expectation}, which is used by the users as a threshold in judging whether the video quality is acceptable or not. In our subjective study, all subjects viewed the videos in a controlled environment and every subject viewed the videos on the same device. Broadly speaking, the video quality expectation $x^*$ can be environment-dependent. For example, viewers tend to have higher expectation for videos shown on a laptop than videos shown on a smartphone. Therefore, in a practical wireless network, $x^*$ may vary across users.
In the following, we denote by $x_u^*$ the video quality expectation of user $u$ and study the following two cases:\\
\noindent\textbf{Case I: Users' video quality expectation is unavailable.} If user $u$ is in the system from $\mathsf{A}_u$ to $\mathsf{D}_u$ and sees video qualities $\left(\mathrm{q}_u(t):~t\in[\mathsf{A}_u,\mathsf{D}_u]\right)$, according to \eqref{eq:f2}, its $2^\mathrm{nd}$-order eCDF is given by
\begin{align}
\label{eq:f2_u}
\textstyle\mathrm{F}^{(2)}\left(x;\mathrm{q}_u\right)=\frac{1}{\mathsf{T}_u}\sum_{t=\mathsf{A}_u}^{\mathsf{D}_u}\left[x-\mathrm{q}_u(t)\right]^+.
\end{align}
If the users' video quality expectation $x^*_u$ is not known \emph{a priori}, we may impose constraints on all $x$ and for all video users. In particular, we consider the following QoE constraints:
\begin{align}
\label{eq:unknown_all}
\mathrm{F}^{(2)}\left(x;\mathrm{q}_u\right)\leq \mathrm{h}(x),~\forall x\in[0,100],~\forall u\in\cup_{t=1}^\infty\mathcal{U}^\mathrm{av}(t),
\end{align}
where $\mathrm{h}(x)$ is a function of $x$.
In practice, we cannot apply constraints on all values of $x\in[0, 100]$. Therefore, we consider a relaxed version of \eqref{eq:unknown_all} as follows:
\begin{align}
\label{eq:unknown}
\mathrm{F}^{(2)}\left(x_i;\mathrm{q}_u\right)\leq \mathrm{h}(x_i),~\forall x_i\in\mathcal{I},~\forall u\in\cup_{t=1}^\infty\mathcal{U}^\mathrm{av}(t).
\end{align}
Here, $\mathcal{I}$ is a discrete set of points on $[0, 100]$. The following property of $2^\mathrm{nd}$-order eCDFs shows that \eqref{eq:unknown} will approximate \eqref{eq:unknown_all} if $\mathcal I$ is dense. Its proof is given in Appendix~\ref{append:2}.
\begin{theorem}
\label{lemm:2}
Let $\bar{\mathrm{h}}(x)$ be the piece-wise linear function that connects the points $\left\{(x_i,\mathrm{h}(x_i)):~\forall x_i\in\mathcal{I}\right\}$. The constraint \eqref{eq:unknown} is equivalent to $\mathrm{F}^{(2)}\left(x;\mathrm{q}_u\right)\leq \bar{\mathrm{h}}(x),~\forall x\in[0,100]$. 
\end{theorem}

\noindent\textbf{Case II: Users' video quality expectation is known.} We also consider the case where users' video quality expectation $x_u^*$ is known or specified by the service provider. For example, users with different viewing devices tend to have different quality expectations. Desktop users usually watch high-definition television programs on large screens and smartphone users usually watch low resolution videos. Thus, we may conduct subjective studies on different devices. Based on the results of the studies, the users' typical video quality expectations $x_u^*$ on each type of device can be deduced. Then, we can categorize video users according to their respective devices and provide differentiated QoE guarantees.

We define a finite set $\mathcal{G}=\{g_1,\dots,g_{|\mathcal{G}|}\}$ that represents different video quality expectations and assume $x^*_u\in\mathcal{G},~\forall u\in\cup_{t=1}^\infty\mathcal{U}^\mathrm{av}(t)$. Let  $\mathcal{U}^\mathrm{av}_j(t)=\{u\in\mathcal{U}^\mathrm{av}(t): x^*_u=g_j\}$ denote the video users whose quality expectation is $g_j$. We consider the following constraints:
\begin{align}
\label{eq:known}
\mathrm{F}^{(2)}(g_j;\mathrm{q}_u)\leq h_j,~\forall g_j\in\mathcal{G},~\forall u\in\cup_{t=1}^\infty\mathcal{U}_j^\mathrm{av}(t),
\end{align}
where $h_j$ is the QoE constraint for users with quality expectation $g_j$.

In sum, the goal of the admission control strategy and the rate adaptation algorithm is to maximize the number of users satisfying constraints \eqref{eq:channelmodel}-\eqref{eq:ratequality}, \eqref{eq:unknown}, or \eqref{eq:known}. In Section~\ref{sec:caseI}, we introduce our rate adaptation algorithm and the admission control strategy when users' quality expectations are unknown. Then, in Section~\ref{sec:case2}, we extend our rate adaptation and admission control algorithms to the case where each user's video quality expectation is known. For ease of reading, a summary of the notation used in this paper is given in Table~\ref{tab:nomenclature}.

\begin{table}
\centering
\caption{NOTATION SUMMARY}
\label{tab:nomenclature}
\begin{tabular}{c|l}
\hline
$\mathsf{A}_u,\mathsf{D}_u$&The arrival and departure time of user $u$\\
\hline
$\mathsf{T}_u$&Time spent by user $u$ in the network\\
\hline
$\mathcal{U}^\mathrm{v}(t)$&Video users at time $t$;\\
\hline
$\mathcal{U}^\mathrm{av}(t)$&Admitted video users at time $t$;\\
\hline
$\mathrm{r}_u(t)$&Data rate of user $u$ at $t$\\
\hline
$\mathcal{C}(t)$&Rate region at $t$\\
\hline
$\mathcal{R}(t)$&The set of available source rates at $t$\\
\hline
$\mathrm{q}_u(t)$&Video quality of user $u$ at $t$\\
\hline
$\mathrm{F}^{(2)}(x;\mathrm{q}_u)$&The second order eCDF of user $u$\\
\hline
$x^*_u$&Video quality expectation of user $u$\\
\hline
$\mathcal I$& Constrained points on eCDFs\\
\hline
$\mathrm{h}(x_i)$& The QoE constraint at $x_i\in\mathcal{I}$\\
\hline
$\mathcal G$& Quality expectations of video users\\
\hline
$\mathcal{U}^\mathrm{av}_j$& Admitted video users with $x^*_u=g_j\in\mathcal{G}$\\
\hline
$h_j$& The QoE constraint for the users in $\mathcal{U}^\mathrm{av}_j$\\
\hline
\end{tabular}
\end{table}
\section{Rate Adaptation and Admission Control with Unknown Video Quality Expectation}
\label{sec:caseI}
If the quality expectations of video users are not known, we apply the same constraint on the second-order eCDFs of all video users. We first propose a rate adaptation algorithm and a corresponding admission control strategy. Then, we evaluate their performance via numerical simulation.
\subsection{Rate Adaptation Algorithm}
\label{sec:rate_caseI}
To clarify the design of our rate adaptation algorithm, we present an off-line problem formulation in which the future channel conditions and admission decisions are assumed to be known. Then, based on the analysis of this offline problem, we propose a new on-line rate adaptation algorithm.

If we consider a finite horizon $T$ and assume that the realizations of channel conditions $\mathcal{C}(1), \dots, \mathcal{C}(T)$ are known, the rate adaptation algorithm should solve the following feasibility problem:
\begin{subequations}
\label{eq:feasibility0}
\begin{eqnarray}
\label{eq:rate_control}
    {\mathrm{find}}&& \mathbf{r}_{1:T}\\
    \mbox{subject to:}&&\mathbf{r}(t)\in\mathcal{C}(t)\cap\mathcal{R}(t),~\forall 1\leq t\leq T\label{eq:channel_20}\\
    && \mathrm{q}_u(t)=\alpha_u(t)\log(\mathrm{r}_u(t))+\beta_u(t),~\forall u\in\mathcal{U}^\mathrm{av}(t),~\forall 1\leq t\leq T\label{eq:ratequalityconstr_20}\\
    && \mathrm{F}^{(2)}\left(x_i;\mathrm{q}_u\right)\leq \mathrm{h}(x_i),~\forall x_i\in\mathcal{I},~\forall u\in\cup_{t=1}^T\mathcal{U}^\mathrm{av}(t).\label{eq:qoe_constr_20}
\end{eqnarray}
\end{subequations}
The constraint \eqref{eq:channel_20} is associated with the achievable rate region \eqref{eq:channelmodel} and the available video source rates in \eqref{eq:min_max}. The constraint \eqref{eq:ratequalityconstr_20} is because of the rate-quality model \eqref{eq:ratequality}. The constraints \eqref{eq:qoe_constr_20} are the QoE constraints \eqref{eq:unknown} that were discussed in Section~\ref{sec:constraints}. For each admitted user, a series of QoE constraints are applied to the $2^\mathrm{nd}$-order eCDF at discrete points in $\mathcal I=\left\{x_1,\cdots,x_{|\mathcal I|}\right\}$. Since the rate-quality function \eqref{eq:ratequalityconstr_20} is concave, according to \cite{Boyd}, the problem \eqref{eq:feasibility0} is equivalent to the following convex optimization problem:
\begin{subequations}
\label{eq:feasibility}
\begin{eqnarray}
\label{eq:rate_control}
    \underset{\begin{subarray}{c}\mathbf{r}_{1:T},~\left(\mathrm{\hat q}_u\right)_{1:T}\end{subarray}}{\mathrm{maximize}}&& 0\\
    \mbox{subject to:}&&\mathbf{r}(t)\in\mathcal{C}(t)\cap\mathcal{R}(t),~\forall 1\leq t\leq T\label{eq:channel_2}\\
    && \mathrm{\hat q}_u(t)\leq\alpha_u(t)\log(\mathrm{r}_u(t))+\beta_u(t),~\forall u\in\mathcal{U}^\mathrm{av}(t),~\forall 1\leq t\leq T\label{eq:ratequalityconstr_2}\\
    && \mathrm{F}^{(2)}\left(x_i;\mathrm{\hat q}_u\right)\leq \mathrm{h}(x_i),~\forall x_i\in\mathcal{I},~\forall u\in\cup_{t=1}^T\mathcal{U}^\mathrm{av}(t),\label{eq:qoe_constr_2}
\end{eqnarray}
\end{subequations}
where $\mathrm{\hat q}_u(t)$ are virtual variables introduced here to make the constraint \eqref{eq:ratequalityconstr_2} convex. Note that the right-hand side of constraint \eqref{eq:ratequalityconstr_2} equals $\mathrm{q}_u(t)$. For any $\mathrm{\hat q}_u(t)$ satisfying \eqref{eq:ratequalityconstr_2}, we have $\mathrm{\hat q}_u(t)\leq \mathrm{q}_u(t)$. Hence, if $\mathrm{\hat q}_u(t)$ satisfies \eqref{eq:qoe_constr_2}, the constraint $\mathrm{F}^{(2)}\left(x_i;\mathrm{q}_u\right)\leq \mathrm{h}(x_i)$ is satisfied as well.

By the definition in \eqref{eq:f2_u}, the $2^\mathrm{nd}$-order eCDF in constraint \eqref{eq:qoe_constr_2} is determined by the entire process $\mathrm{\hat q}_u(\mathsf{A}_u), \dots, \mathrm{\hat q}_u(\mathsf{D}_u)$. Due to the constraints \eqref{eq:channel_2} and \eqref{eq:ratequalityconstr_2}, $\mathrm{\hat q}_u(t)$ depends on the rate $\mathrm{r}_u(t)$ and thus also depends on the rate region $\mathcal{C}(t)$. Therefore, the solution of \eqref{eq:feasibility} depends on the entire process $\mathcal{C}(1), \dots, \mathcal{C}(T)$. In practice, the future channel conditions are unavailable to the rate-adaptation algorithm. In the following, we transform problem \eqref{eq:feasibility} to a simpler form that inspires our online rate adaptation algorithm.

Since \eqref{eq:feasibility} is a convex problem, if it is feasible, there exists a set of Lagrange multipliers $\lambda^*_{u,i}\geq0$ for the constraints in \eqref{eq:qoe_constr_2} such that a solution of \eqref{eq:feasibility} can be obtained by solving the following problem (see \cite{Convex}):
\begin{subequations}
\label{eq:lagrangian}
\begin{eqnarray}
\underset{\begin{subarray}{c}\mathbf{r}_{1:T},~\left(\mathrm{\hat q}_u\right)_{1:T}\end{subarray}}{\mathrm{minimize}}&& \sum_{u\in\cup_{t=1}^T\mathcal{U}^\mathrm{av}(t)}\sum_{x_i\in\mathcal{I}}\lambda^*_{u,i}\left(\mathrm{F}^{(2)}\left(x_i;\mathrm{\hat q}_u\right)-\mathrm{h}(x_i)\right)\label{eq:lagrangian_objective}\\
\mbox{subject to:}&&\mathbf{r}(t)\in\mathcal{C}(t)\cap\mathcal{R}(t),~\forall 1\leq t\leq T\label{eq:channel_3}\\
&& \mathrm{\hat q}_u(t)\leq\alpha_u(t)\log(\mathrm{r}_u(t))+\beta_u(t),~\forall u\in\mathcal{U}^\mathrm{av}(t),~\forall 1\leq t\leq T.\label{eq:ratequalityconstr_3}
\end{eqnarray}
\end{subequations}
If we define a function $\mathrm{s}_{u,i}(t)$ as
\begin{align}
\label{eq:s}
    \mathrm{s}_{u,i}(t)=
\begin{mycases}
\frac{1}{\mathsf{T}_u}\left(\left[ x_i-\mathrm{\hat q}_u(t)\right]^+-\mathrm{h}(x_i)\right)& \text{if } \mathsf{A}_u\leq t\leq\mathsf{D}_u,\\
0& \text{otherwise},
\end{mycases}
\end{align}
the term  $\mathrm{F}^{(2)}\left(x_i;\mathrm{\hat q}_u\right)-\mathrm{h}(x_i)$ in \eqref{eq:lagrangian_objective} can be rewritten as
\begin{equation}
\label{eq:rewritef2}
\begin{aligned}
&\mathrm{F}^{(2)}\left(x_i;\mathrm{\hat q}_u\right)-\mathrm{h}(x_i)\\
=&\textstyle{\sum_{t=\mathsf{A}_u}^{\mathsf{D}_u}\frac{1}{\mathsf{T}_u}\left(\left[ x_i-\mathrm{\hat q}_u(t)\right]^+-\mathrm{h}(x_i)\right)}\\
=&\textstyle{\sum_{t=1}^{T}\mathrm{s}_{u,i}(t)}.
\end{aligned}
\end{equation}
Thus, $\mathrm{s}_{u,i}(t)$ indicates to what extent the variable $\mathrm{\hat q}_u(t)$ violates the constraint $\mathrm{F}^{(2)}\left(x_i;\mathrm{\hat q}_u\right)\leq\mathrm{h}(x_i)$ in each slot. Substituting \eqref{eq:rewritef2} in \eqref{eq:lagrangian_objective} and changing the order of summation, 
the optimization in \eqref{eq:lagrangian} becomes
\begin{equation}
\label{eq:greedy}
\begin{aligned}
\underset{\begin{subarray}{c}\mathbf{r}_{1:T},~\left(\mathrm{\hat q}_u\right)_{1:T}\end{subarray}}{\mathrm{minimize}}&\quad \sum_{t=1}^T\left(\sum_{u\in\mathcal{U}^\mathrm{av}(t)}\sum_{x_i\in\mathcal{I}}\lambda^*_{u,i}\mathrm{s}_{u,i}(t)\right)\\
\mbox{subject to:}&\quad\mathbf{r}(t)\in\mathcal{C}(t)\cap\mathcal{R}(t),~\forall 1\leq t\leq T\\
&\quad \mathrm{\hat q}_u(t)\leq\alpha_u(t)\log(\mathrm{r}_u(t))+\beta_u(t),~\forall u\in\mathcal{U}^\mathrm{av}(t),~\forall 1\leq t\leq T.
\end{aligned}
\end{equation}
Note that, except for the Lagrange multipliers, the optimization in \eqref{eq:greedy} does not involve variables that depend on the entire process $\{\mathrm{\hat q}_u(t): 1\leq t\leq T\}$. Thus, \eqref{eq:greedy} can be solved by minimizing the weighted sum $\sum_{u\in\mathcal{U}^\mathrm{av}(t)}\sum_{x_i\in\mathcal{I}}\lambda^*_{u,i}\mathrm{s}_{u,i}(t)$ in every slot. That is, if it is possible to estimate the Lagrange multiplier $\lambda^*_{u,i}$, then \eqref{eq:feasibility} can be solved by greedily choosing the rate vector $\mathbf{r}(t)$ at each time slot as the solution of the following problem:
\begin{equation}
\label{eq:greedy_2}
\begin{aligned}
\underset{\mathbf{r}(t),~\mathrm{\hat q}_u(t)}{\mathrm{minimize}}&\quad \sum_{u\in\mathcal{U}^\mathrm{av}(t)}\sum_{x_i\in\mathcal{I}}\lambda^*_{u,i}\mathrm{s}_{u,i}(t)\\
\mbox{subject to:}&\quad\mathbf{r}(t)\in\mathcal{C}(t)\cap\mathcal{R}(t)\\
&\quad \mathrm{\hat q}_u(t)\leq\alpha_u(t)\log(\mathrm{r}_u(t))+\beta_u(t),~\forall u\in\mathcal{U}^\mathrm{av}(t).
\end{aligned}
\end{equation}

We introduce a method to approximate the Lagrange multiplier $\lambda^*_{u,i}$. We know that the Lagrange multiplier $\lambda^*_{u,i}$ indicates the difficulty in satisfying the constraint $\mathrm{F}_u^{(2)}(x_i;\mathrm{\hat q}_u)\leq \mathrm{h}(x_i)$ \cite{Boyd}. Inspired by prior work in \cite{Stolyar2005} and \cite{Neelybook}, we employ a virtual queue to capture this difficulty. For each admitted user $u$ and each $x_i\in\mathcal{I}_u$, define the virtual queue as
\begin{align}
\label{eq:queue}
\mathrm{v}_{u,i}(t)=
\begin{mycases}
\left[\mathrm{v}_{u,i}(t-1)+\mathrm{s}_{u,i}(t)\right]^+&\text{if }\mathsf{A}_u\leq t\leq\mathsf{D}_u,\\
0&\text{otherwise}.
\end{mycases}
\end{align}
From \eqref{eq:rewritef2} it follows that, if the summation of $\mathrm{s}_{u,i}(t)$ is large, then it is difficult to satisfy the constraint $\mathrm{F}_u^{(2)}(x_i;\mathrm{\hat q}_u)\leq \mathrm{h}(x_i)$. The virtual queue captures the cumulative summation of $\mathrm{s}_{u,i}(t)$ up to slot $t$. Hence, the virtual queue reflects the level of difficulty in satisfying $\mathrm{F}_u^{(2)}(x_i;\mathrm{\hat q}_u)\leq \mathrm{h}(x_i)$. Actually, for the special case where user set $\mathcal{U}^\mathrm{av}(t)$ is fixed for all $t$, it can be proved that the virtual queue asymptotically approaches $\lambda^*_{u,i}$ as $T\rightarrow\infty$ \cite{Stolyar2005}. Hence, we replace the Lagrange multipliers in \eqref{eq:greedy_2} with virtual queue $\mathrm{v}_{u,i}(t)$ and our online rate adaptation algorithm is summarized in Algorithm \ref{alg:online_rate}. In every slot, we maximize the weighted sum of $\mathrm{s}_{u,i}(t)$, where the weight is given by $\mathrm{v}_{u,i}(t-1)$. Thus users with larger virtual queues tend to be allocated more network resources. This helps users satisfy their QoE constraints.

Next, we introduce an admission control policy that is combined with our rate-adaptation algorithm to further improve performance.

\begin{algorithm}[h]
\caption{Online algorithm for video data rate adaptation}
\label{alg:online_rate}
\begin{algorithmic}[1]
\For{$t = 1 \to \infty$}
\State Choose rate vector $\mathbf{r}(t)$ that solves the problem
\begin{equation}
\label{eq:greedy_3}
\begin{aligned}
\underset{\mathbf{r}(t),~\mathrm{\hat q}_u(t)}{\mathrm{minimize}}&\quad \sum_{u\in\mathcal{U}^\mathrm{av}(t)}\sum_{x_i\in\mathcal{I}}\mathrm{v}_{u,i}(t-1)\mathrm{s}_{u,i}(t)\\
\mbox{subject to:}&\quad\mathbf{r}(t)\in\mathcal{C}(t)\cap\mathcal{R}(t)\\
&\quad \mathrm{\hat q}_u(t)\leq\alpha_u(t)\log(\mathrm{r}_u(t))+\beta_u(t),~\forall u\in\mathcal{U}^\mathrm{av}(t),
\end{aligned}
\end{equation}
\quad\  where $\mathrm{s}_{u,i}(t)=\frac{1}{\mathsf{T}_u}\left(\left[ x_i-\mathrm{\hat q}_u(t)\right]^+-\mathrm{h}(x_i)\right)$.
\State For $\forall u\in\mathcal{U}^\mathrm{av}(t), \forall x_i\in\mathcal{I}$, update virtual queues with \[\mathrm{v}_{u,i}(t)=[\mathrm{v}_{u,i}(t-1)+\mathrm{s}_{u,i}(t)]^+.\]
\EndFor
\end{algorithmic}
\end{algorithm}

\subsection{Admission Control Strategy}
\label{sec:admisson_case1}
Since a video stream typically has high data rate and thus consumes a large amount of resources, the arrival and departure of a single video user can have a significant impact on other video users' QoE. Our admission control strategy is designed to identify and block those video users who may consume excessive network resources. As has been discussed in {Algorithm~\ref{alg:online_rate}}, resource allocation in each slot is determined by the solution of the optimization problem \eqref{eq:greedy_3}. Therefore, it is possible to estimate the QoE of a newly arrived user by solving \eqref{eq:greedy_3} as if the user had already been admitted. Based on this idea, we propose a threshold-based admission control strategy, which is summarized in {Algorithm~\ref{alg:online_ac_case1}}. For each newly arrived video user ${\bar u}$, we first estimate its video quality ${\bar q}$ by solving the optimization problem \eqref{eq:greedy_4}, which is similar to the optimization problem \eqref{eq:greedy_3}. Then, we compare ${\bar q}$ with a threshold $\theta$. If ${\bar q}$ is larger than $\theta$, it is admitted to the network. Otherwise, it is rejected.
\renewcommand{\algorithmicrequire}{\textbf{Inputs:}}
\begin{algorithm}[h]
\caption{Admission control when video quality expectation is not known.}
\label{alg:online_ac_case1}
\begin{algorithmic}[1]
\Require Threshold $\theta$, admitted users $\mathcal{U}^\mathrm{av}(t-1)$, new user ${\bar u}$
\State Initialize video user set $\mathcal{U}^\mathrm{av+}\leftarrow\mathcal{U}^\mathrm{av}(t-1)\cap\{{\bar u}\}$
\State Estimate mean rate-quality parameters for $\forall u\in\mathcal{U}^\mathrm{av+}$:
    \begin{equation}
    \begin{aligned}
{\hat\alpha}_u&\leftarrow1/{\mathsf{T}_u}\textstyle{\sum_{t=\mathsf{A}_u}^{\mathsf{D}_u}\alpha_u(t)},\\     {\hat\beta}_u&\leftarrow1/{\mathsf{T}_u}\textstyle{\sum_{t=\mathsf{A}_u}^{\mathsf{D}_u}\beta_u(t)}.
    \end{aligned}
    \end{equation}
\State Initialize virtual queue for a new user ${\bar u}$:
    \begin{equation}\textstyle{
    \mathrm{v}_{{\bar u},i}(t-1)\leftarrow\frac{1}{|\mathcal{U}^\mathrm{av}(t-1)|}\sum_{u\in\mathcal{U}^\mathrm{av}(t-1)}\mathrm{v}_{u,i}(t-1),~\forall i\in\mathcal{I}}
    \end{equation}
\State Define variables $\mathbf{\hat r}=\left({\hat r}_u:~u\in\mathcal{U}^\mathrm{av+}\right)$, $\mathbf{\hat q}=\left({\hat q}_u:~u\in\mathcal{U}^\mathrm{av+}\right)$. Also define variables ${\hat s}_{u,i}=\frac{1}{\mathsf{T}_u}\left(\left[ x_i-{\hat q}_u\right]^+-\mathrm{h}(x_i)\right)$.
Find the solution $\mathbf{r}^*=\left(r^*_u:~u\in\mathcal{U}^\mathrm{av+}\right)$ of the optimization problem
\begin{equation}
\label{eq:greedy_4}
\begin{aligned}
\underset{\mathbf{\hat r},\mathbf{\hat q}}{\mathrm{minimize}}&\quad \textstyle{\sum_{u\in\mathcal{U}^\mathrm{av+}}\sum_{x_i\in\mathcal{I}}}\mathrm{v}_{u,i}(t-1){\hat s}_{u,i}\\
\mbox{subject to:}&\quad\mathbf{\hat r}\in\mathcal{C}\cap\mathcal{R},\\
&\quad {\hat q}_u\leq{\hat \alpha}_u\log({\hat r}_u)+{\hat \beta}_u,~\forall u\in\mathcal{U}^\mathrm{av+},
\end{aligned}
\end{equation}
where the sets $\mathcal{C}=\{\mathbf{r}:\mathbb{E}[\mathrm{C}_t(\mathbf{r})\leq 0]\}$ and $
\textstyle
\mathcal{R}=\Pi_{u\in\mathcal{U}^\mathrm{av+}}\left[\frac{1}{\mathsf{T}_u}\sum_{t=\mathsf{A}_u}^{\mathsf{D}_u}\mathrm{r}^\mathrm{min}_u(t),\frac{1}{\mathsf{T}_u}\sum_{t=\mathsf{A}_u}^{\mathsf{D}_u}\mathrm{r}^\mathrm{max}_u(t)\right].
$
\State Estimate the video quality delivered to new user via
\begin{equation}
{\bar q}={\hat\alpha}_{{\bar u}}\log\left(r^*_{{\bar u}}\right)+{\hat\beta}_{{\bar u}}.
\end{equation}
\State If ${\bar q}>\theta$, admit the new user; otherwise, reject it.
\end{algorithmic}
\end{algorithm}

The optimization problem \eqref{eq:greedy_4} is different from \eqref{eq:greedy_3} in the following three aspects. First, to predict the long-term QoE of users, we replace the instantaneous rate-quality parameters $\alpha_u(t)$ and $\beta_u(t)$ in \eqref{eq:greedy_3} with the average rate-quality parameters ${\hat \alpha}_u$ and ${\hat \beta}_u$ (see the second step in {Algorithm~\ref{alg:online_ac_case1}}):
\begin{equation}
    \label{eq:rq_est}
    \begin{aligned}
{\hat\alpha}_u&={\textstyle\frac{1}{\mathsf{T}_u}{\sum_{t=\mathsf{A}_u}^{\mathsf{D}_u}\alpha_u(t)},}\\     {\hat\beta}_u&={\textstyle\frac{1}{\mathsf{T}_u}{\sum_{t=\mathsf{A}_u}^{\mathsf{D}_u}\beta_u(t)}.}
    \end{aligned}
\end{equation}
Second, we replace the instantaneous rate region $\mathcal{C}(t)$ with a rate region estimated by
\begin{equation}
\label{eq:c_est}
\mathcal{C}=\{\mathbf{r}:\mathbb{E}[\mathrm{C}_t(\mathbf{r})]\leq 0\}.
\end{equation} Similarly, we replace the set of available video source rate $\mathcal{R}(t)$ by \begin{equation}
\label{eq:r_est}\textstyle
\mathcal{R}=\Pi_{u\in\mathcal{U}^\mathrm{av}(t-1)\cup\{\bar u\}}\left[\frac{1}{\mathsf{T}_u}\sum_{t=\mathsf{A}_u}^{\mathsf{D}_u}\mathrm{r}^\mathrm{min}_u(t),\frac{1}{\mathsf{T}_u}\sum_{t=\mathsf{A}_u}^{\mathsf{D}_u}\mathrm{r}^\mathrm{max}_u(t)\right].
\end{equation}
Third, for a newly arrived user ${\bar u}$, we initialize its virtual queue with the average virtual queues of the existing users (see the third step in {Algorithm~\ref{alg:online_ac_case1}}), i.e.,
\begin{equation}
\label{eq:queue_est}
    \mathrm{v}_{{\bar u},i}(t-1)\leftarrow\frac{1}{|\mathcal{U}^\mathrm{av}(t-1)|}\sum_{u\in\mathcal{U}^\mathrm{av}(t-1)}\mathrm{v}_{u,i}(t-1),~\forall i\in\mathcal{I}.
\end{equation}

In \eqref{eq:rq_est}, the rate-quality parameters $\alpha_u(\cdot)$ and $\beta_u(\cdot)$ are needed. For stored video streaming systems, the videos are pre-encoded. Thus, we assume the rate-quality parameters for the entire video stream are known. Also, the rate region $\mathcal{C}$ in \eqref{eq:c_est} can be estimated using the time-average of the previous channel conditions.
For example, in TDMA systems, we have $\mathrm{C}_t(\mathbf{r})=\sum_{u\in\mathcal{U}^\mathrm{av}(t)}\frac{{r}_u}{\mathsf{P}_u(t)}+\sum_{{u'}\in\mathcal{U}^\mathrm{p}(t)}\frac{\mathsf{R}_{u'}(t)}{\mathsf{P}_{u'}(t)}-1$ (see Section~\ref{sec:channelmodel}).
Thus, we can estimate $\mathbb{E}\left[\frac{1}{\mathsf{P}_u}\right]$ and $\mathbb{E}\left[\sum_{{u'}\in\mathcal{U}^\mathrm{p}(t)}\frac{\mathsf{R}_{u'}(t)}{\mathsf{P}_{u'}(t)}\right]$ using the previous observations of the peak rate $\mathsf{P}_u$ and the high-priority users' data rate $\mathsf{R}_u$. The estimated rate region is therefore $\mathcal{C}=\left\{\mathbf{r}: \sum_{u\in\mathcal{U}^\mathrm{av}(t-1)\cup\{\bar u\}}\mathbb{E}\left[\frac{1}{\mathsf{P}_u}\right]r_u\leq 1-\mathbb{E}\left[\sum_{{u'}\in\mathcal{U}^\mathrm{p}(t)}\frac{\mathsf{R}_{u'}(t)}{\mathsf{P}_{u'}(t)}\right]\right\}$.

As was discussed in Section~\ref{sec:rate_caseI}, the virtual queue $\mathrm{v}_{u,i}(t)$ captures the difficulty for an admitted video user to satisfy the QoE constraints. Thus, users with large virtual queues tend to be allocated more network resources. In other words, the virtual queues drive the priorities in resource allocation. Because it is difficult to estimate the length of a virtual queue before a new user is admitted, we simply initialize the virtual queue of the newly arrived user with the average virtual queues of all existing video users. In this way, we actually estimate the video quality ${\bar q}$ when an average priority is assigned to the new user. Next, we introduce an approach to optimize the admission threshold $\theta$ in {Algorithm~\ref{alg:online_ac_case1}}.

\subsection{Online Algorithm for Threshold Optimization}
\begin{figure}[!h]
\centering
\includegraphics[width=0.6\columnwidth]{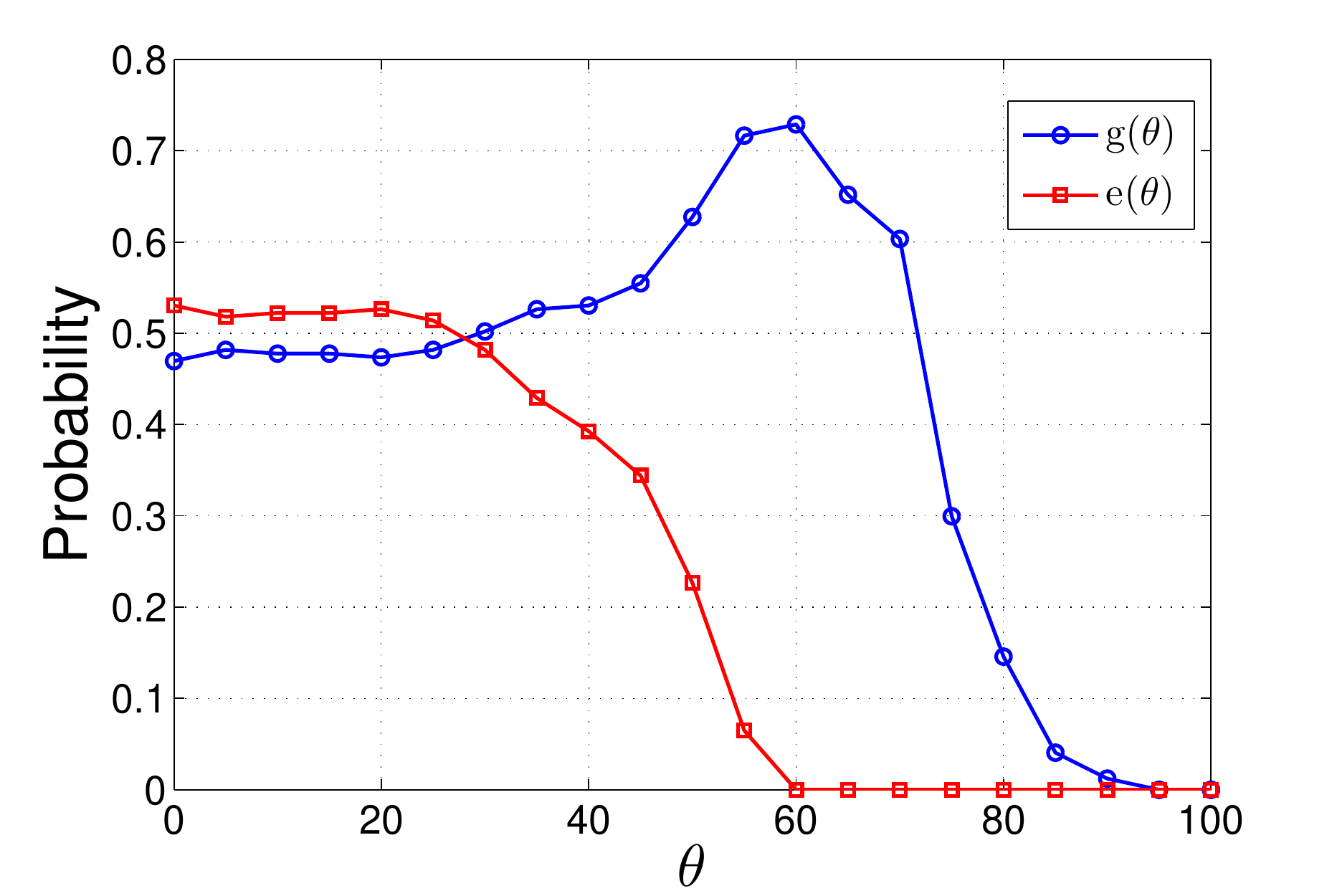}
\caption{The two plots show (i) the percentage of video users who satisfy the QoE constraints and (ii) the percentage of video users who are admitted into the network but do not satisfy the QoE constraints under different admission control thresholds.}
\label{fig:experiment3}
\end{figure}

Denote by $\mathrm{g}(\theta)$ the probability that a video user's QoE constraints are satisfied when the threshold is $\theta$. Also, denote by $\mathrm{e}(\theta)$ the probability that a video user is admitted into the network but its QoE constraints are not satisfied. Our goal is to find the threshold $\theta^*$ that maximizes $\mathrm{g}(\theta)$. We have conducted extensive simulations under different channel conditions and QoE constraints. From all the simulation results, we observed that the optimal threshold $\theta^*$ maximizes $\mathrm{g}(\theta)$ if
\begin{align}
        \label{eq:observation}
        \begin{mycases}
        \mathrm{e}(\theta)>0, & \forall \theta< \theta^*\\
        \mathrm{e}(\theta)=0, & \forall \theta\geq \theta^*
        \end{mycases}.
        \end{align}
This means that $\mathrm{g}(\theta)$ is maximized if $\theta$ is just large enough to make all the admitted users satisfy the QoE constraints. As an example, using the same simulation configurations that are detailed later in Section~\ref{sec:sim_rate_case1}, we simulated and plotted the functions $\mathrm{e}(\theta)$ and $\mathrm{g}(\theta)$ in Fig.~\ref{fig:experiment3}. It is seen that $\theta^*=60$ satisfies \eqref{eq:observation} and $\mathrm{g}(\theta)$ is also maximized at $\theta^*$. Therefore, to find the optimal threshold $\theta^*$, it is sufficient to find a threshold that satisfies \eqref{eq:observation}.

We propose an iterative algorithm that automatically adjusts the threshold to $\theta^*$. This is summarized in {Algorithm~\ref{alg:online_threshold_case1}}. In each iteration, the algorithm observes the $2^\mathrm{nd}$-order eCDFs of $L$ video users who have been admitted into the network since the end of the last iteration. Then the algorithm updates the threshold via
\begin{align}
    \theta^{n+1}=\theta^{n}+\epsilon^{n}y^{n},
\end{align}
where $\theta^{n}$ denote the admission control threshold in the $n^\mathrm{th}$ iteration. The value $y^n\in\{-1,1\}$ determines whether to increase or to decrease the threshold. The quantity $\epsilon^n>0$ is an updating step size. If the algorithm observes a video user whose $2^\mathrm{nd}$-order eCDF violates the QoE constraints, then it is probable that $\mathrm{e}(\theta^n)>0$ and $\theta^n<\theta^*$. Therefore, the algorithm increases the threshold by setting $y^n=1$. Otherwise, if all the $L$ video users satisfy the constraints, the threshold is possibly larger than $\theta^*$. Thus the algorithm decreases the threshold by setting $y^n=-1$. The updating step size is
\begin{align}
\label{eq:step_size_case1}
\epsilon^n=\epsilon^0/m,
\end{align}
where $m$ counts the number of sign changes in the series $\{y^1, \dots, y^n\}$ (see step 8) and $\epsilon^0$ is the initial step-size. Here, $m$ is introduced to accelerate the convergence of the algorithm. The reason is as follows: If $\theta^n$ is far from $\theta^*$, the sign of $y^n$ does not change frequently and $m$ increases slowly. Thus, the step-size $\epsilon^n$ stays large and $\theta^n$ moves towards $\theta^*$ quickly. When $\theta^n$ is moved to a small neighborhood of $\theta^*$, the sign of $y^n$ changes frequently and thus $m$ increases rapidly. Therefore, the step-size $\epsilon^n$ decreases to zero quickly, which makes $\theta^n$ converge.
\renewcommand{\algorithmicrequire}{\textbf{Inputs:}}
\begin{algorithm}[h]
\caption{The threshold optimization algorithm when the video quality expectation is unknown.}
\label{alg:online_threshold_case1}
\begin{algorithmic}[1]
\Require $L=100$, initial threshold $\theta^0=0$, initial step-size $\epsilon^0=10$, and initial counter $m=1$
\For {$n=1\rightarrow\infty$}
\State Observe the $2^\mathrm{nd}$-order eCDFs of $L$ admitted video users.
\If {there exits a user that does not satisfy the QoE constraints}
    \State $y^n\leftarrow 1$
\Else
    \State $y^n\leftarrow -1$
\EndIf
\State If $y^n\neq y^{n-1}$, $m\leftarrow m+1$.
\State Update threshold with
    \begin{align}
    \theta^{n+1}=\theta^n+\epsilon^ny^n,
    \end{align}
    \quad\ \ where $\epsilon^n={\epsilon^0}/{m}$.
\EndFor
\end{algorithmic}
\end{algorithm}

In the following, we analyze the convergence of Algorithm~\ref{alg:online_threshold_case1}.
Based on our observations from the simulations, we make the following assumption:
\newtheorem{assumption}{\bf Assumption}
\begin{assumption}
\label{asump:e_k}
The function $\mathrm{e}(\theta)$ is a continuous function and is strictly decreasing on $[0,~\theta^*]$.
\end{assumption}
We define $\mathrm{p}^L(\theta)$ as the probability that all the $L$ admitted video users in an iteration satisfy the QoE constraints. Since increasing the threshold $\theta$ would block more users and thus reserve more network resources to the admitted users, we assume that $\mathrm{p}^L(\theta)$ is a continuously increasing function of $\theta$. Also, according to Assumption~\ref{asump:e_k}, when $\theta>\theta^*$, all the admitted users satisfy the QoE constraints and thus $\mathrm{p}^L(\theta)=1$. Thus, we have the following assumption on $\mathrm{p}^L(\theta)$:
\begin{assumption}
\label{asump:p_k}
The function $\mathrm{p}^L(\theta)$ is a continuous and increasing function of $\theta$. For $\forall \theta>\theta^*$, we have $\mathrm{p}^L(\theta)=1$. Furthermore, assume that there exists a constant $M>0$ such that $|\mathrm{p}^L(\theta)-\mathrm{p}^L(\theta')|\geq M|\theta-\theta'|$ for all $\theta'< \theta$ and $\mathrm{p}^L(\theta)<1$.
\end{assumption}

The following theorem assures that if $L$ is sufficiently large, $\theta^n$ converges to an arbitrarily small neighborhood of $\theta^*$ as $n\rightarrow\infty$. Its proof is given in Appendix~\ref{append:1}.
\begin{theorem}
\label{Tho:1}
Let $\delta>0$ be an arbitrarily small number. If Assumptions~\ref{asump:e_k} and \ref{asump:p_k} are satisfied and $L\geq\frac{-\log2}{\log(1-\mathrm{e}(\theta^*-\delta))}$, then $\theta^n$ converges as $n\rightarrow\infty$ and  $\lim_{n\rightarrow\infty}\theta^n\in[\theta^*-\delta,\theta^*]$.
\end{theorem}
\subsection{Simulation Results}
\label{sec:sim_rate_case1}
Below, we evaluate our rate-adaptation algorithm and the admission control strategy via numerical simulations. We assume the duration of a time slot is $\Delta T=1\text{ second}$. The high-priority users' arrivals follow a Poisson process with average arrival rate of $\frac{1}{20}\text{ users/second}$. The time spent by a high-priority user in the network is exponentially distributed with a mean value of $200\text{ seconds}$. Video users also arrival as a Poisson process with a average arrival rate of $\frac{1}{20}\text{ users/second}$.
Since video streams are typically more than tens of seconds long, we assume the time spent by a video user in the network is at least 40 seconds. In particular, for all video users, we set $\mathsf{T}_u=\max\{\mathsf{T'}_u,40\}$, where $\mathsf{T'}_u$ is exponentially distributed with a mean value of $200\text{ seconds}$.

To simulate variations of the rate-quality characteristics in each video stream, we assume the rate-quality parameters $(\alpha_u(t),\beta_u(t))$ of each slot are independently sampled from the rate-quality parameters in the video database \cite{chaoICASSP}. We assume the minimum and maximum available data rate for video users in \eqref{eq:min_max} are $\mathrm{r}^\mathrm{min}_u(t)=302\text{ kbps}$ and $\mathrm{r}^\mathrm{max}_u(t)=6412\text{ kbps}$, which are the minimum and maximum rate of the videos in the database \cite{chaoICASSP}.
For high-priority users, the downloading data rate $\mathsf{R}_u$ is assumed to be uniformly distributed in $[100, 300]\text{ kbps}$.

We assume the wireless system is a TDMA system. The rate region $\mathcal{C}(t)$ is that introduced in Section~\ref{sec:channelmodel}. We model the peak transmission rate $\mathsf{P}_u(t)$ as the product of two independent random variables, i.e., $\mathsf{P}_u(t)=\mathsf{P}_u^\mathrm{avg}\times\mathsf{P}_u^*(t)$. The random variable $\mathsf{P}_u^\mathrm{avg}$ is employed to simulate the heterogeneity of channel condition across users and remains constant during a user's sojourn. We assume that $\mathsf{P}_u^\mathrm{avg}$ is uniformly distributed on $[1250\gamma, 3750\gamma]\text{ kbps}$, where the parameter $\gamma$ is used to scale the channel capacity in our simulations. The random variable $\mathsf{P}_u^*(t)$ is employed to simulate channel variation across time slots. We assume that $\{\mathsf{P}_u^*(t):t\in\mathbb{N}^+\}$ is an i.i.d. process with $\mathsf{P}_u^*(t)$ being uniformly distributed on $[0.5,1.5]$. In our simulations, we set $\mathcal{I}=\{30, 40, 50, 60,70\}$. Correspondingly, for $x_i=$ 30, 40, 50, 60, and 70, we let the constraints $\mathrm{h}(x_i)=$ 0.7, 1.0, 3.0, 7.0 and 15.0, respectively.
\begin{figure}[!h]
\centering
\subfigure[Proposed rate-adaptation algorithm]{\includegraphics[width=0.6\columnwidth]{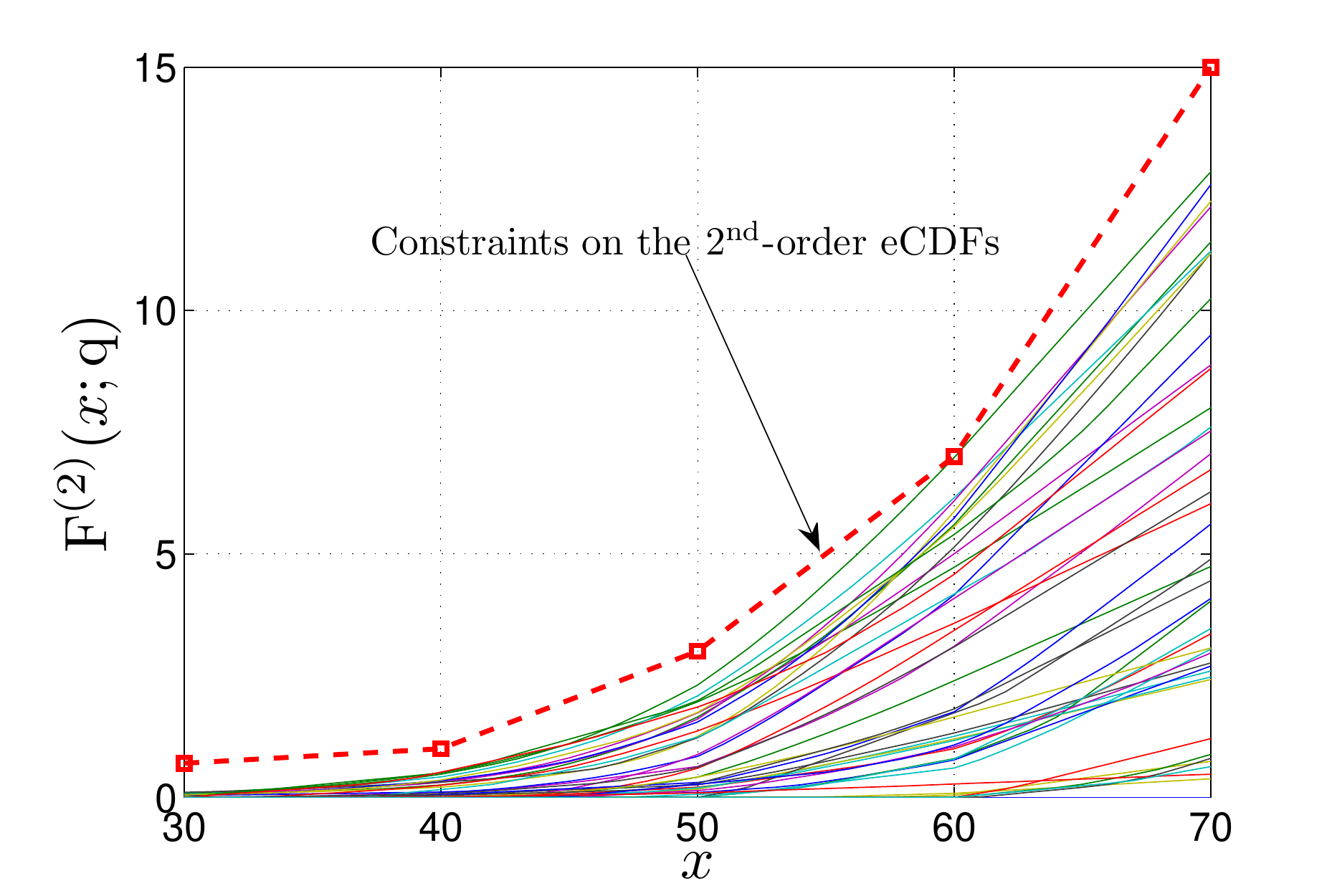}
\label{fig:exp_1_alpha_10}}
\subfigure[Averge-quality maximized rate adaptation]{\includegraphics[width=0.6\columnwidth]{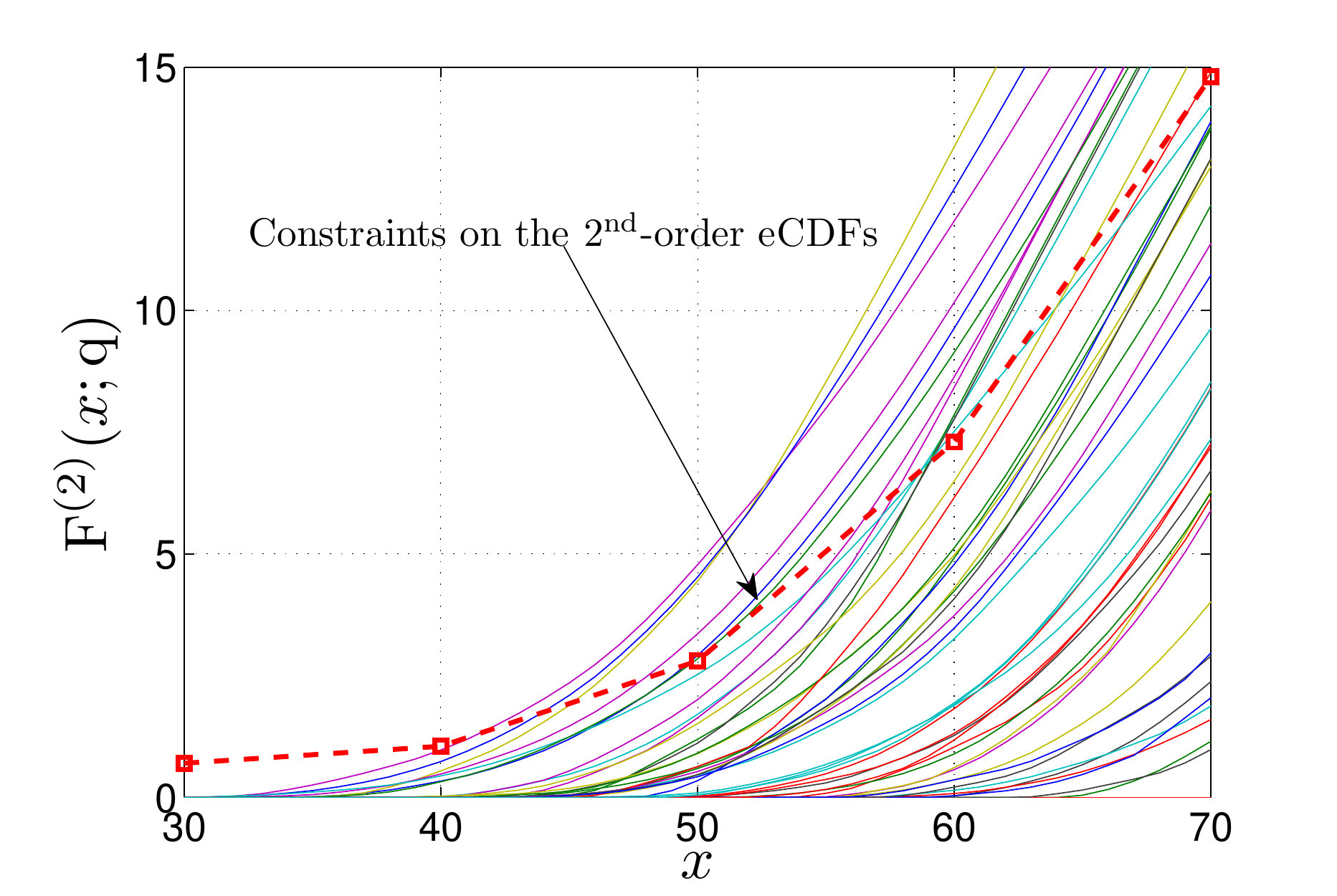}
\label{fig:exp_1_max_q}
}
\caption{Simulation results of rate-adaptation algorithms when admission control is not applied. \subref{fig:exp_1_alpha_10} The $2^\mathrm{nd}$-order eCDFs of the video users when the proposed rate-adaptation is used. \subref{fig:exp_1_max_q} The $2^\mathrm{nd}$-order eCDFs of the video users when the rate vector is adapted to maximize the sum of users' video qualities.}
\label{fig:experiment1}
\end{figure}

We first evaluate the performance of the rate-adaptation algorithm when admission control is not applied. We set the scaling parameter $\gamma=12$ and simulate Algorithm \ref{alg:online_rate} until 100 users have arrived and departed the network. We plot the $2^\mathrm{nd}$-order eCDFs of the video users in Fig.~\ref{fig:exp_1_alpha_10}. It may be seen that, using  Algorithm \ref{alg:online_rate}, the $2^\mathrm{nd}$-order eCDFs of the video users all satisfy the constraints. By comparison, if we adapt the rate vector to maximize the sum of the average-quality of all users\footnote{This is achieved by maximizing $\sum_{u\in\mathcal{U}^\mathrm{av}(t)}\mathrm{q}_u(t)/\mathsf{T}_u$ in each slot.}, the QoE constraint is violated by many users (see Fig.~\ref{fig:exp_1_max_q}).

\begin{figure}[!h]
\centering
\subfigure[]{\includegraphics[width=0.6\columnwidth]{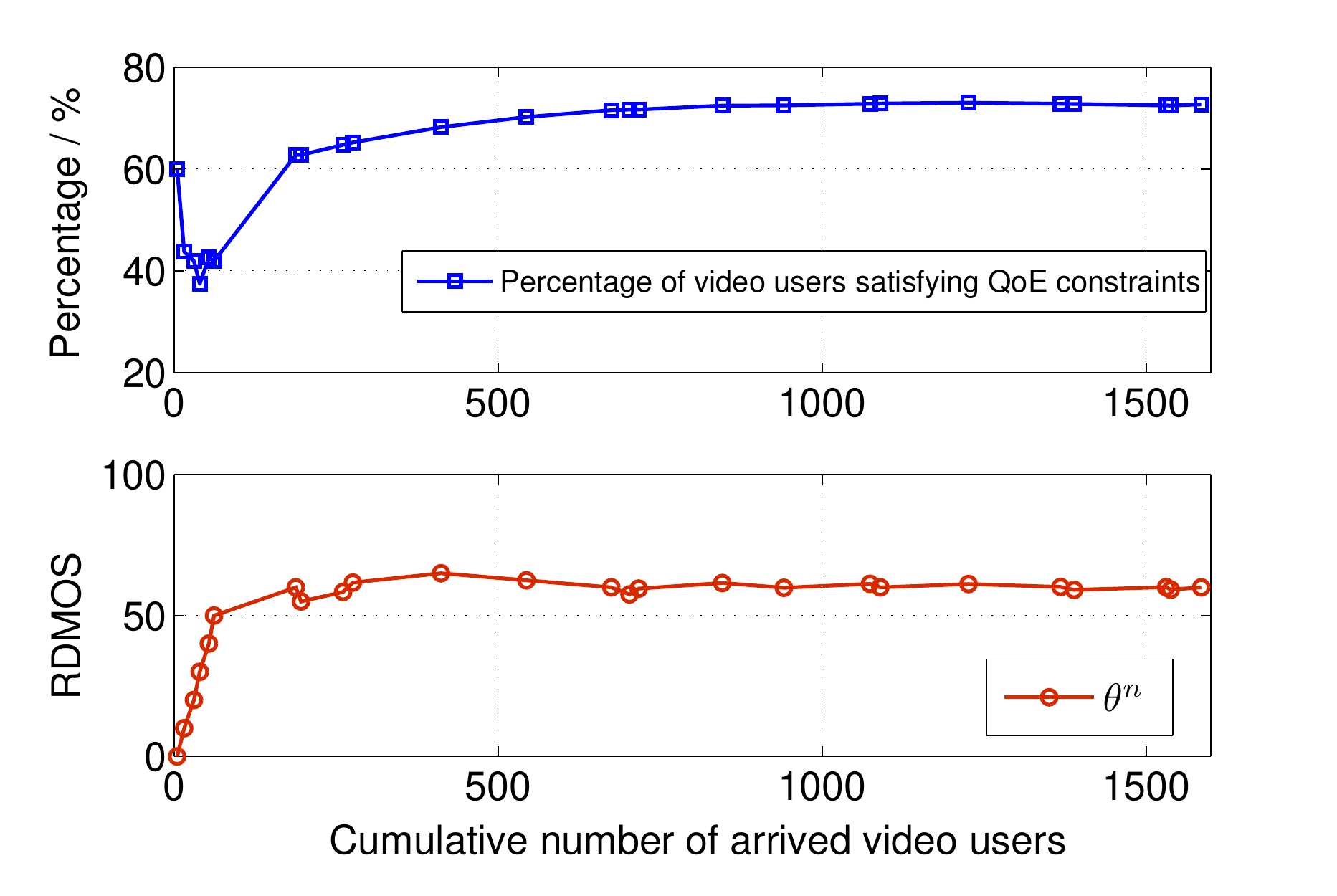}
\label{fig:experiment3_online}
}
\subfigure[]{\includegraphics[width=0.6\columnwidth]{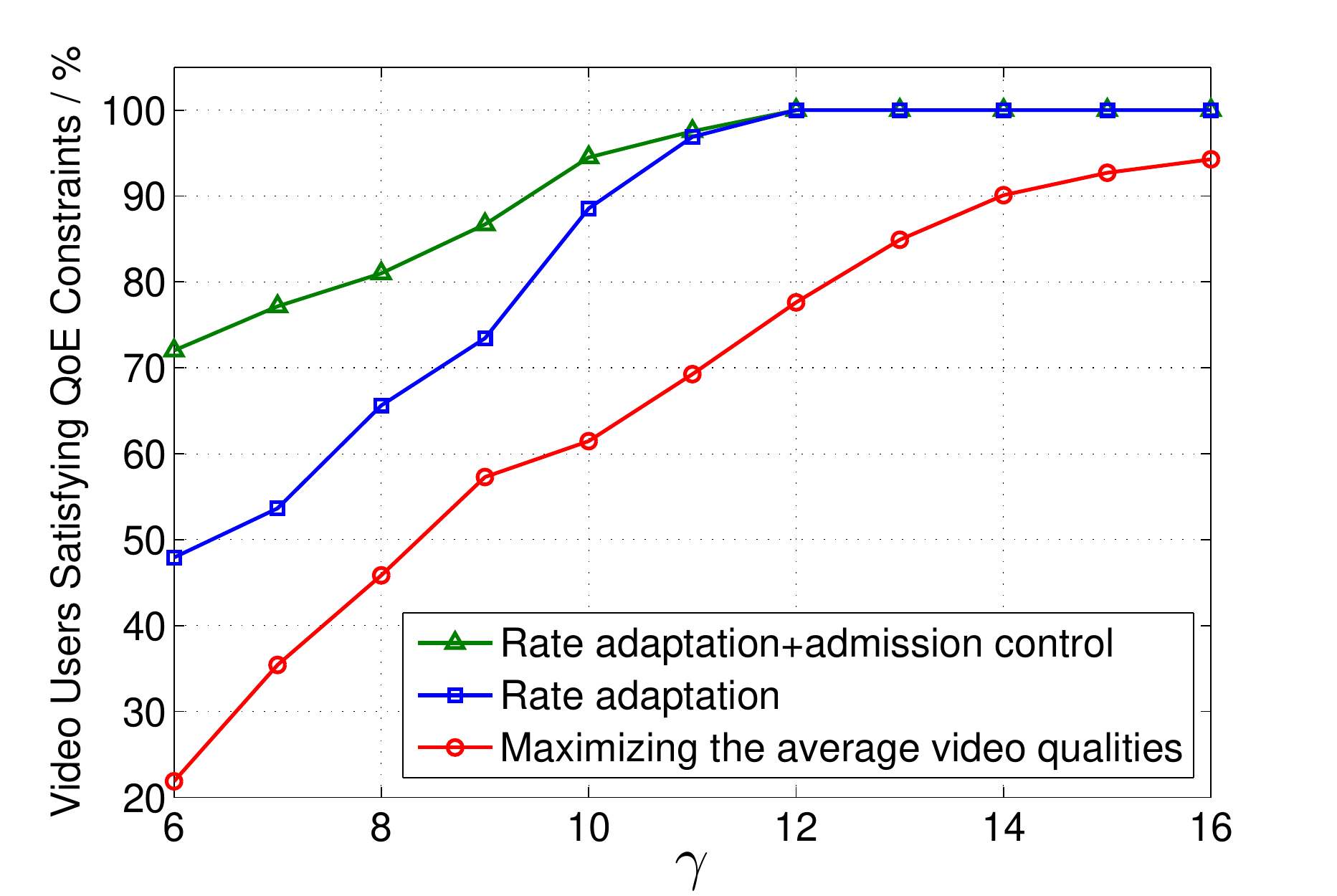}
\label{fig:experiment3_scale_channel}
}
\caption{\subref{fig:experiment3_online} The performance of the proposed admission control strategy when the scaling parameter is $\gamma=6$. \subref{fig:experiment3_scale_channel} Simulation results of the proposed algorithms under different channel scaling parameters. Each data point on the figure is obtained by simulating 2000 video user arrivals.}
\label{fig:experiment3_performance}
\end{figure}

Next, we evaluate the proposed admission control strategy. In Fig.~\ref{fig:experiment3_performance}\subref{fig:experiment3_online}, we fix the channel scaling parameter to be $\gamma=6$ and plot the threshold $\theta^n$ at every iteration of the online threshold optimization algorithm. Recall that the optimal threshold is $\theta^*=60$ (see Fig.~\ref{fig:experiment3}). Fig. \ref{fig:experiment3_online} shows that $\theta^n$ converges to $\theta^*$ after 200 video user arrivals. We have assumed that the average arrival rate of video users is $1/20$ users/seconds. Thus, 200 video user arrivals require about $200\times20 \text{ seconds}=1.1\text{ hours}$. Since our goal is to optimize the performance of the network in the long run, this convergence speed is acceptable.

We scale the channel scaling parameter from $\gamma=6$ to $\gamma=16$.
The percentage of video users whose video qualities satisfy the QoE constraints is shown in Fig.~\ref{fig:experiment3_performance}\subref{fig:experiment3_scale_channel}. When compared with the average-quality-maximized rate-adaptation algorithm, the percentage of video users who satisfy the QoE constraints is improved significantly even if no admission control is applied. The admission control policy further improves the performance especially when the channel condition is poor. For example, at $\gamma=6$, the proposed algorithms satisfy the QoE constraints of 70\% of the video users while the average-quality-maximized rate-adaptation algorithm only satisfies the constraints of 20\% of the video users. At a moderate channel condition of $\gamma=12$, about 77\% of the video users satisfy the QoE constraints when the average-quality maximizing algorithm is applied. The proposed algorithms achieve the same performance at $\gamma=7.5$, reducing the consumption of resources by $(12-7.5)/12=38\%$.

\section{Rate Adaptation and Admission Control with Known Video Quality Expectation}
\label{sec:case2}
In this section, we extend the rate adaptation algorithm and the admission control strategy to the case where the video quality expectation of each user is known. We first explain the extended algorithms and then evaluate their performance via simulation.

\subsection{The Extended Rate Adaptation and Admission Control Algorithms}
\label{sec:rate_case2}
In Section~\ref{sec:constraints}, we defined the finite set $\mathcal{G}=\{g_1,\dots,g_{|\mathcal{G}|}\}$ to represent different video quality expectations among video users. In the following, we call users with $x^*_u=g_j\in\mathcal{G}$ the Type-$j$ users. Each Type-$j$ video user need only satisfy one QoE constraint, i.e.,
\begin{align}
\label{eq:known_2}
\mathrm{F}^{(2)}(g_j;\mathrm{q}_u)\leq h_j.
\end{align}
Thus, we extend the rate adaptation method in Algorithm~\ref{alg:online_rate} by maintaining one virtual queue for each user. In particular, the virtual queue of a Type-$j$ user $u$ is defined as
\begin{align}
\label{eq:queue_case2}
\mathrm{v}_{u}(t)=
\begin{mycases}
\left[\mathrm{v}_{u}(t-1)+\mathrm{s}_u(t)\right]^+&\text{if }\mathsf{A}_u\leq t\leq\mathsf{D}_u,\\
0&\text{otherwise}
\end{mycases}
\end{align}
where
\begin{equation}
\mathrm{s}_u(t)=
\begin{mycases}
\frac{1}{\mathsf{T}_u}\left(\left[ g_j-\mathrm{\hat q}_u(t)\right]^+-h_j\right)&\text{if }\mathsf{A}_u\leq t\leq\mathsf{D}_u,\\
0&\text{otherwise}
\end{mycases}.
\end{equation}
In each slot, the rate vector $\mathbf{r}(t)$ is adapted by solving
\begin{equation}
\label{eq:greedy_case2}
\begin{aligned}
\underset{\mathbf{r}(t),\mathrm{\hat q}_u(t)}{\mathrm{minimize}}&\quad \sum_{u\in\mathcal{U}^\mathrm{av}(t)}\mathrm{v}_u(t-1)\mathrm{s}_u(t)\\
\mbox{subject to:}&\quad\mathbf{r}(t)\in\mathcal{C}(t)\cap\mathcal{R}(t),\\
&\quad \mathrm{\hat q}_u(t)\leq\alpha_u(t)\log(\mathrm{r}_u(t))+\beta_u(t),~\forall u\in\mathcal{U}^\mathrm{av}(t).
\end{aligned}
\end{equation}

For admission control, we extend Algorithm~\ref{alg:online_ac_case1} by applying different thresholds to different types of users. In particular, for a newly arrived Type-$j$ video user ${\bar u}$, we initialize its virtual queue by averaging the virtual queues of all existing video users i.e.,
\begin{equation}
\label{eq:queue_est_case2}
\mathrm{v}_{{\bar u}}(t-1)\leftarrow\frac{1}{|\mathcal{U}^\mathrm{av}(t-1)|}\sum_{u\in\mathcal{U}^\mathrm{av}(t-1)}\mathrm{v}_u(t-1).
\end{equation}
Letting $\mathcal{U}^\mathrm{av+}=\mathcal{U}^\mathrm{av}(t)\cup\{{\bar u}\}$, we define variables $\mathbf{\hat r}=\left({\hat r}_u:~u\in\mathcal{U}^\mathrm{av+}\right)$, $\mathbf{\hat q}=\left({\hat q}_u:~u\in\mathcal{U}^\mathrm{av+}\right)$, and ${\hat s}_u=\frac{1}{\mathsf{T}_u}\left(\left[ g_j-{\hat q}_u\right]^+-h_j\right)$. We then find the solution $\mathbf{r^*}=\left(r^*_u:~u\in\mathcal{U}^\mathrm{av+}\right)$ of the following problem:
\begin{equation}
\label{eq:greedy_6}
\begin{aligned}
\underset{\mathbf{\hat r},\mathbf{\hat q}}{\mathrm{minimize}}&\quad {\sum_{u\in\mathcal{U}^\mathrm{av+}}}\mathrm{v}_u(t-1){\hat s}_u\\
\mbox{subject to:}&\quad\mathbf{\hat r}\in\mathcal{C}\cap\mathcal{R},\\
&\quad {\hat q}_u\leq{\hat\alpha}_u\log({\hat r}_u)+{\hat\beta}_u,~\forall u\in\mathcal{U}^\mathrm{av+}
\end{aligned}
\end{equation}
where ${\hat\alpha}_u$, ${\hat\beta}_u$, $\mathcal{C}$, and $\mathcal{R}$ are given by \eqref{eq:rq_est}, \eqref{eq:c_est}, and \eqref{eq:r_est}, respectively.
Finally, we estimate the video quality of the new user by ${\bar q}={\hat\alpha}_{{\bar u}}\log(r^*_{{\bar u}})+{\hat\beta}_{{\bar u}}$ and compare ${\bar q}$ with a threshold $\theta_j$ to make the admission decision.

Next, we discuss how to optimize the threshold $\theta_j$ for Type-$j$ users.
\subsection{The Extended Threshold Optimization Algorithm}
\begin{figure}[!h]
    \centering
    \subfigure[$\mathrm{e}_1(\boldsymbol{\theta})$]{\includegraphics[width=0.31\columnwidth]{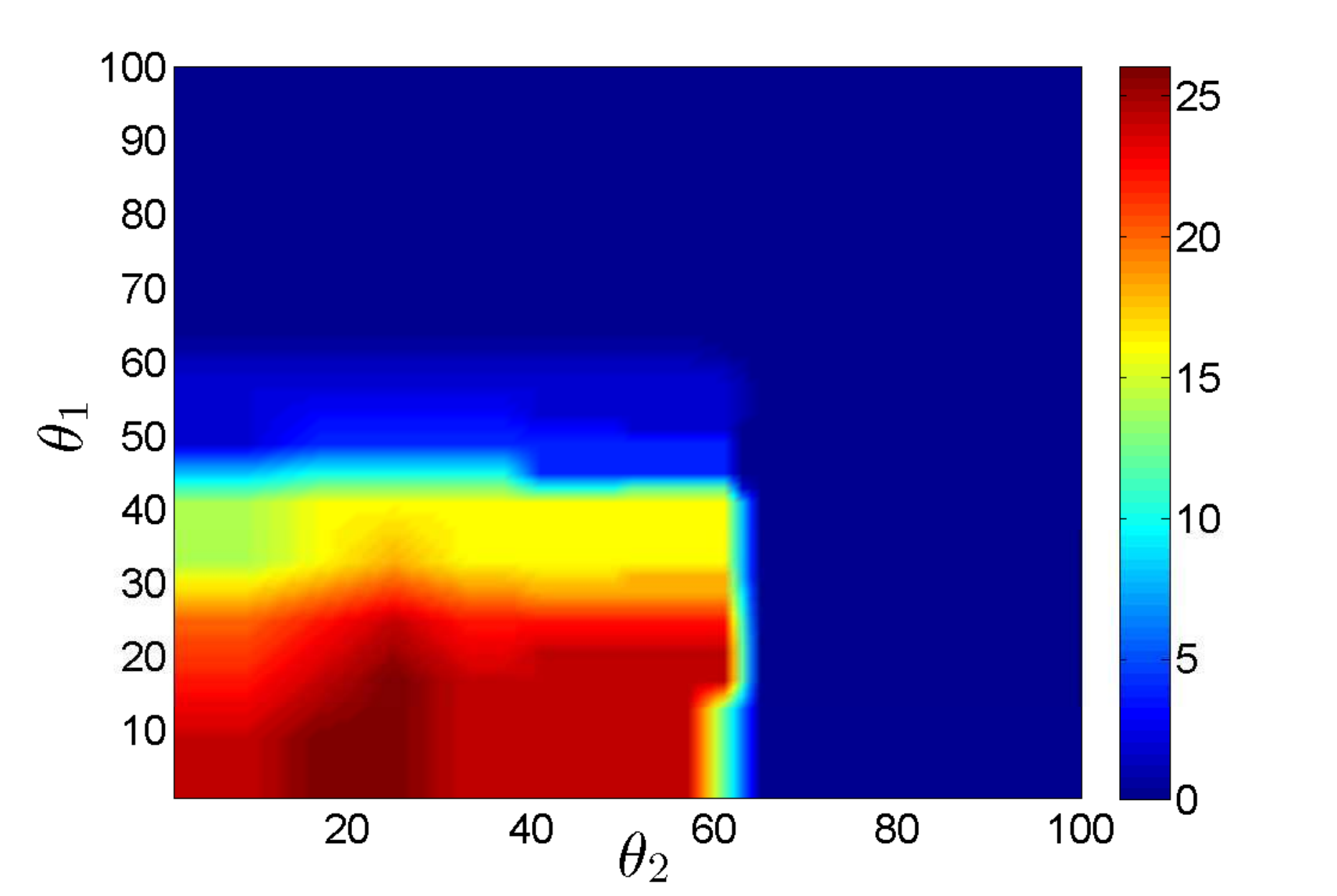}
    \label{fig:e1}}
    \hfil
    \subfigure[$\mathrm{e}_2(\boldsymbol{\theta})$]{\includegraphics[width=0.31\columnwidth]{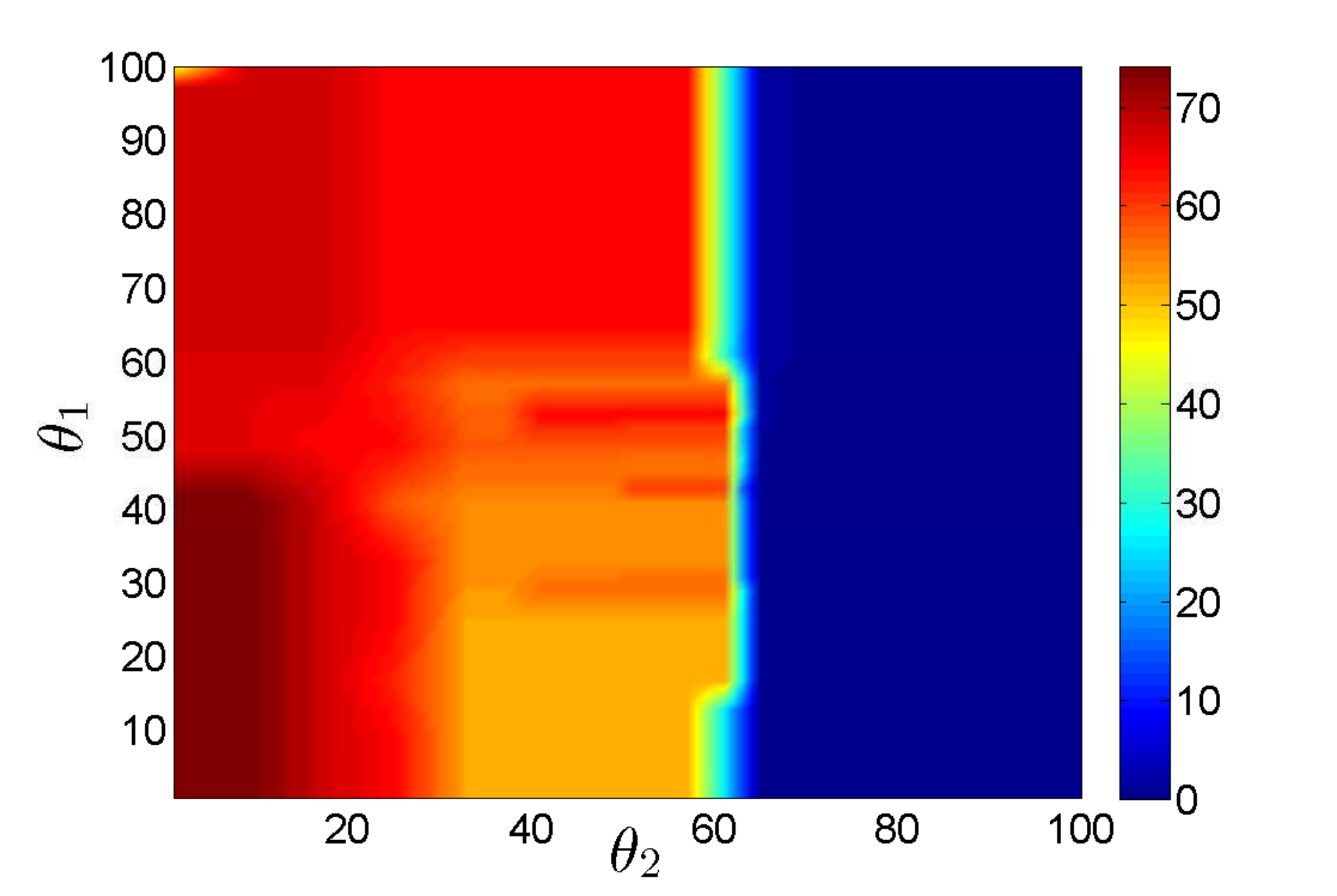}
    \label{fig:e2}}
    \hfil
    \subfigure[$\mathrm{g}(\boldsymbol{\theta})$]{\includegraphics[width=0.31\columnwidth]{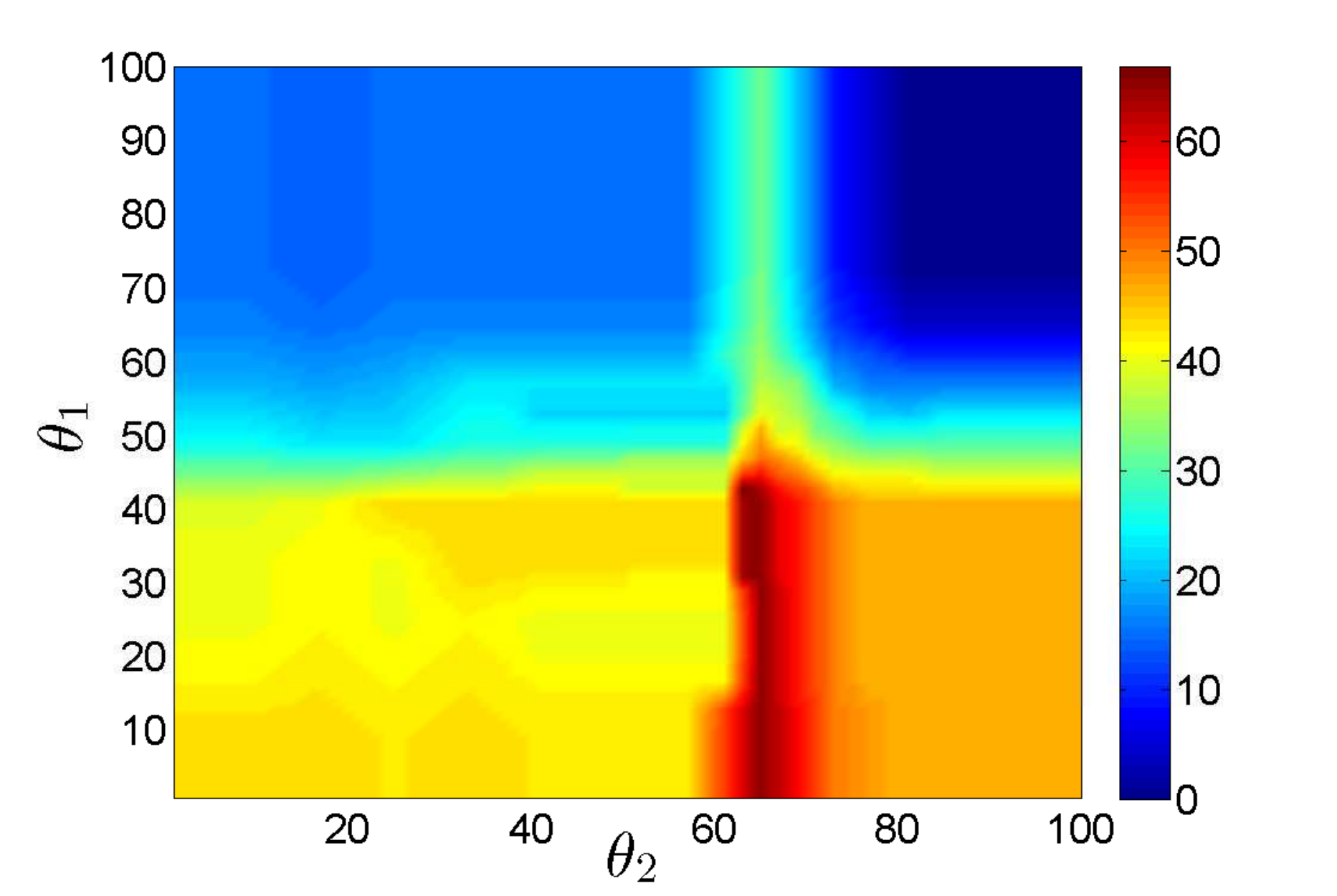}
    \label{fig:g}}
    \hfil
    \caption{ \subref{fig:e1} The probability of admitted Type-1 video users whose QoE constraints are violated. \subref{fig:e2} The probability of admitted Type-2 video users whose QoE constraints are violated. \subref{fig:g} The probability of video users whose QoE constraints are satisfied. All results are shown in percentages. We assume these two types of user have $x_u^*=$ $40$ and $60$, respectively. The QoE constraint in \eqref{eq:known_2} is assumed to be $h_1=h_2=1$. The channel scaling parameter is set as $\gamma=6$.}
    \label{fig:experiment4_example}
\end{figure}

Denote by $\boldsymbol{\theta}=(\theta_1,\dots,\theta_{|\mathcal{G}|})$ the vector of thresholds for all types of video users. Define $\mathrm{g}(\boldsymbol{\theta})$ to be the probability that a video user satisfies the QoE constraints when the threshold vector is $\boldsymbol{\theta}$. Also, define $\mathrm{e}_j(\boldsymbol{\theta})$ as the probability that a Type-$j$ video user's QoE constraint is not satisfied. To determine the optimal threshold vector $\boldsymbol{\theta}$ that maximizes $\mathrm{g}(\boldsymbol{\theta})$, we ran simulations under a variety of relevant channel conditions and QoE constraints. We found that a threshold vector $\boldsymbol{\theta}^*=\left(\theta^*_1,\dots,\theta_{|\mathcal G|}^*\right)$ maximizes $\mathrm{g}(\boldsymbol{\theta})$ if
\begin{align}
\label{eq:observation_2}
\begin{mycases}
\mathrm{e}_j(\boldsymbol{\theta})>0,&\forall~\boldsymbol{\theta}\prec\boldsymbol{\theta}^*\\
\mathrm{e}_j(\boldsymbol{\theta})=0,&\forall~\boldsymbol{\theta}\succeq\boldsymbol{\theta}^*.
\end{mycases},~\forall 1\leq j\leq|\mathcal G|.
\end{align}
Here, the partial order $\boldsymbol{\theta}\prec\boldsymbol{\theta}^*$ indicates that $\boldsymbol{\theta}\neq\boldsymbol{\theta}^*$ and $\theta_j\leq \theta^*_j,~\forall j$. The partial order $\boldsymbol{\theta}\succeq\boldsymbol{\theta}^*$ indicates that $\theta_j\geq \theta^*_j,~\forall j$. The condition in \eqref{eq:observation_2} means that if $\boldsymbol{\theta}^*$ is an optimal threshold vector and we increase all entries of $\boldsymbol{\theta}^*$, the QoE constraints of all the admitted users can still be satisfied. Conversely, if we decrease all the entries of $\boldsymbol{\theta}^*$, the QoE constraints of all types of users will be violated with a non-zero probability. To illustrate this, we considered two types of video users who arrive to the network with equal probability and simulated the function $\mathrm{e}_1(\boldsymbol{\theta})$, $\mathrm{e}_2(\boldsymbol{\theta})$, and $\mathrm{g}_1(\boldsymbol{\theta})$ using the same setting as in Section~\ref{sec:sim_rate_case1}. From Fig.~\ref{fig:e1} and Fig.~\ref{fig:e2}, it can be seen that the $\boldsymbol\theta$s in $[0,42]\times[63,64]$ satisfy the condition \eqref{eq:observation_2} because they lie on the boundary of the region where $\{\mathrm{e}_1(\boldsymbol{\theta})>0$ and $\mathrm{e}_2(\boldsymbol{\theta})>0\}$. From Fig.~\ref{fig:g}, it is seen that the function $\mathrm{g}(\boldsymbol{\theta})$ is also maximized on the region $[0,42]\times[63,64]$.

We devise an iterative algorithm to find the threshold vector $\boldsymbol{\theta}^*$ satisfying \eqref{eq:observation_2}. Denote by $\boldsymbol{\theta}^n$ the threshold vector at the $n^\mathrm{th}$ iteration, the algorithm observes the $2^\mathrm{nd}$-order eCDFs of $L$ admitted video users and updates the threshold vector using
\begin{align}
\label{eq:updating_2}
    \boldsymbol{\theta}^{n+1}=\boldsymbol{\theta}^n+\mathrm{diag}(\boldsymbol{\epsilon}^n)\mathbf{y}^n,
\end{align}
where $\mathrm{diag}(\boldsymbol{\epsilon}^n)$ is the diagonal matrix with diagonal entries being $\epsilon_1,\dots,\epsilon_{|\mathcal G|}$. In \eqref{eq:updating_2}, ${\mathbf y}^n=\left(y_1^n,\dots,y_{|\mathcal G|}^n\right)$ is a $|\mathcal G|$-dimensional vector where $y_j^n\in\{-1,1\}$ is the updating direction for $\theta_j$. The vector $\boldsymbol\epsilon=\left(\epsilon_1^n,\dots,\epsilon_{|\mathcal G|}^n\right)$ is the corresponding update step-size. Among the $L$ video users, if a Type-$j$ video user's QoE constraint is not satisfied, the algorithm sets $y_j^n=1$. Otherwise, if all the Type-$j$ video users' QoE constraints are satisfied, the algorithm sets $y_j^n=-1$. The step-size $\epsilon_j^n$ is given by $\epsilon_j^n=\epsilon_j^0/m_j$, where $m_j$ counts the sign changes in $\left\{y_j^1,\dots,y_j^n\right\}$ and $\epsilon_j^0$ is the initial step-size. Next, we show the performance of our rate adaptation algorithm and the admission control strategy via simulation.


\subsection{Simulation Results}

\begin{figure}[!h]
\centering
\includegraphics[width=0.65\columnwidth]{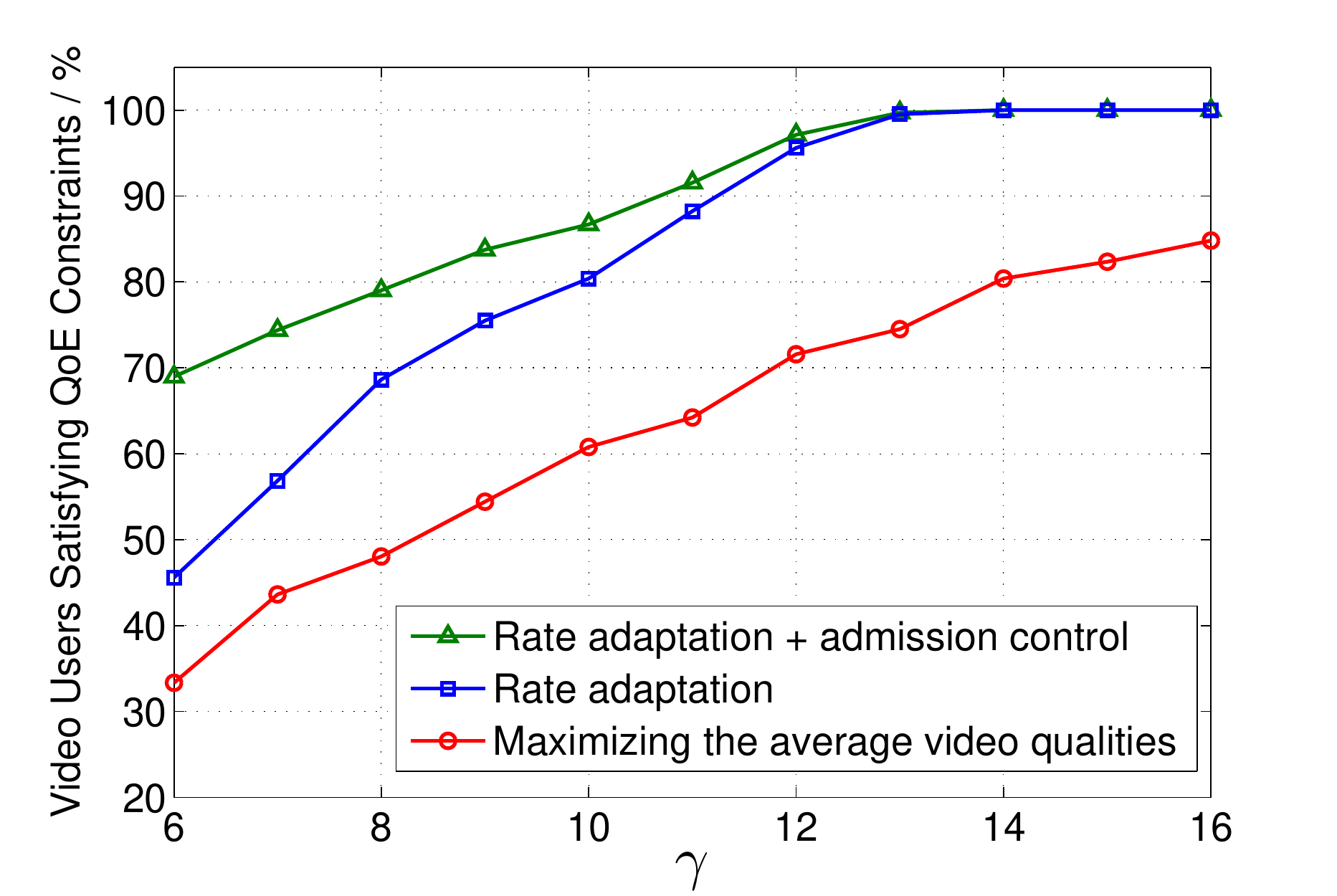}
\caption{Simulation results of the proposed admission control policy under different channel scaling parameters.}
\label{fig:experiment4_scale_alpha}
\end{figure}

In our simulations, we assume that there are two types of video users. Both types of video users arrive as a Poisson process with arrival rate $1/40\text{ users/second}$. We assume that Type-1 users have $x_u^*=40$ while Type-2 users have $x_u^*=60$. We also set $h_1=h_2=1$. In Fig.~\ref{fig:experiment4_scale_alpha}, we plot the percentage of video users whose video qualities satisfy the QoE constraint \eqref{eq:known}. It can be seen that, for all tested channel scaling parameters, our rate adaptation algorithm outperforms the average-quality maximizing algorithm. The percentage of video users who satisfy the QoE constraints improved significantly even when the admission control strategy was not applied. At $\gamma=12$, about 71\% of video users satisfied the QoE constraints when the average-quality maximizing algorithm was applied. The proposed algorithms achieve the same performance at $\gamma=6.5$. Thus, the proposed algorithms reduce the consumption of resources by $(12-6.5)/12=46\%$.

\begin{figure}[!h]
    \centering
    \subfigure[Convergence performance of the admission control strategy]{\includegraphics[width=.45\columnwidth]{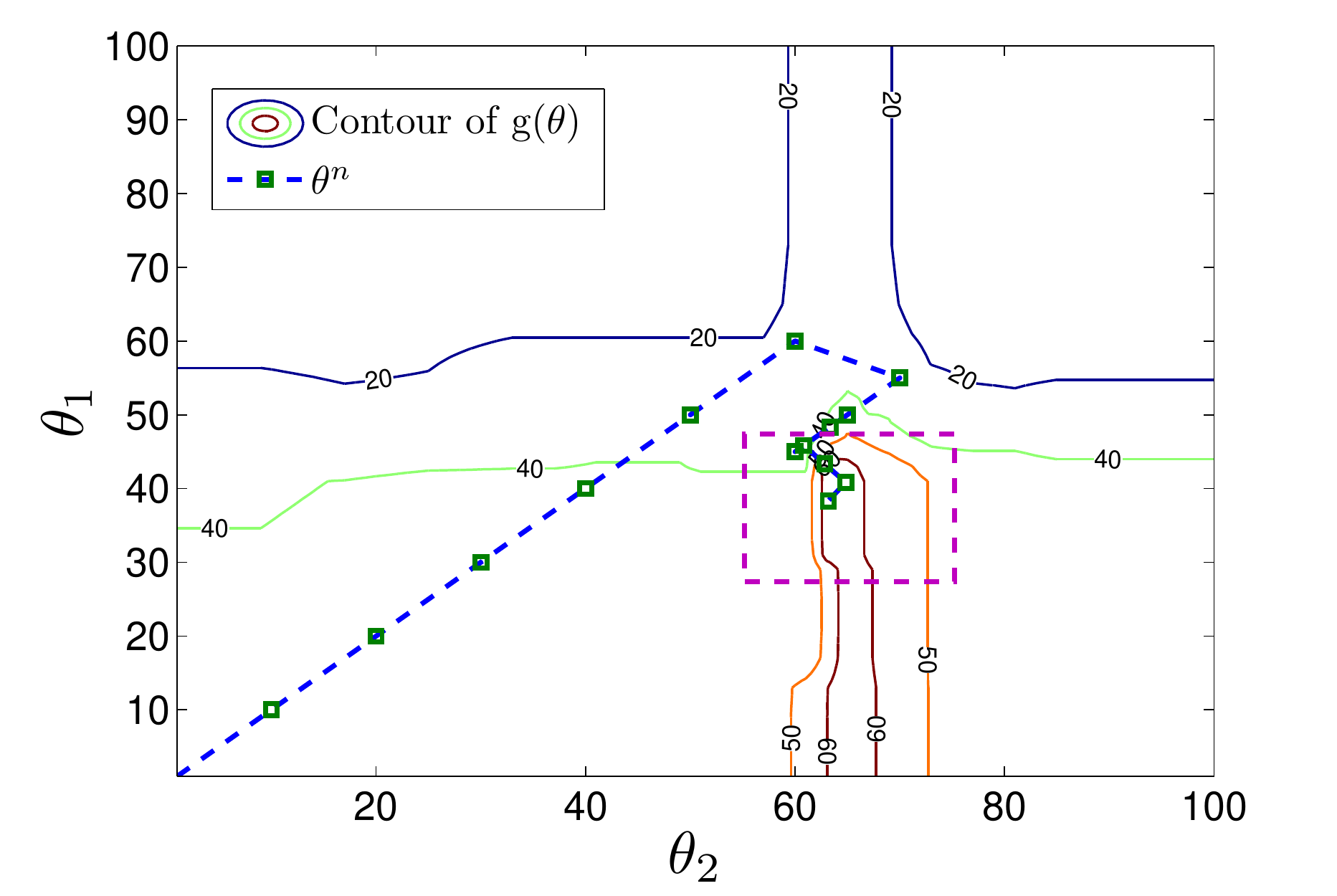}
    \label{fig:experiment4_large}}
    \hfil
    \subfigure[A zoom-in view]{\includegraphics[width=.45\columnwidth]{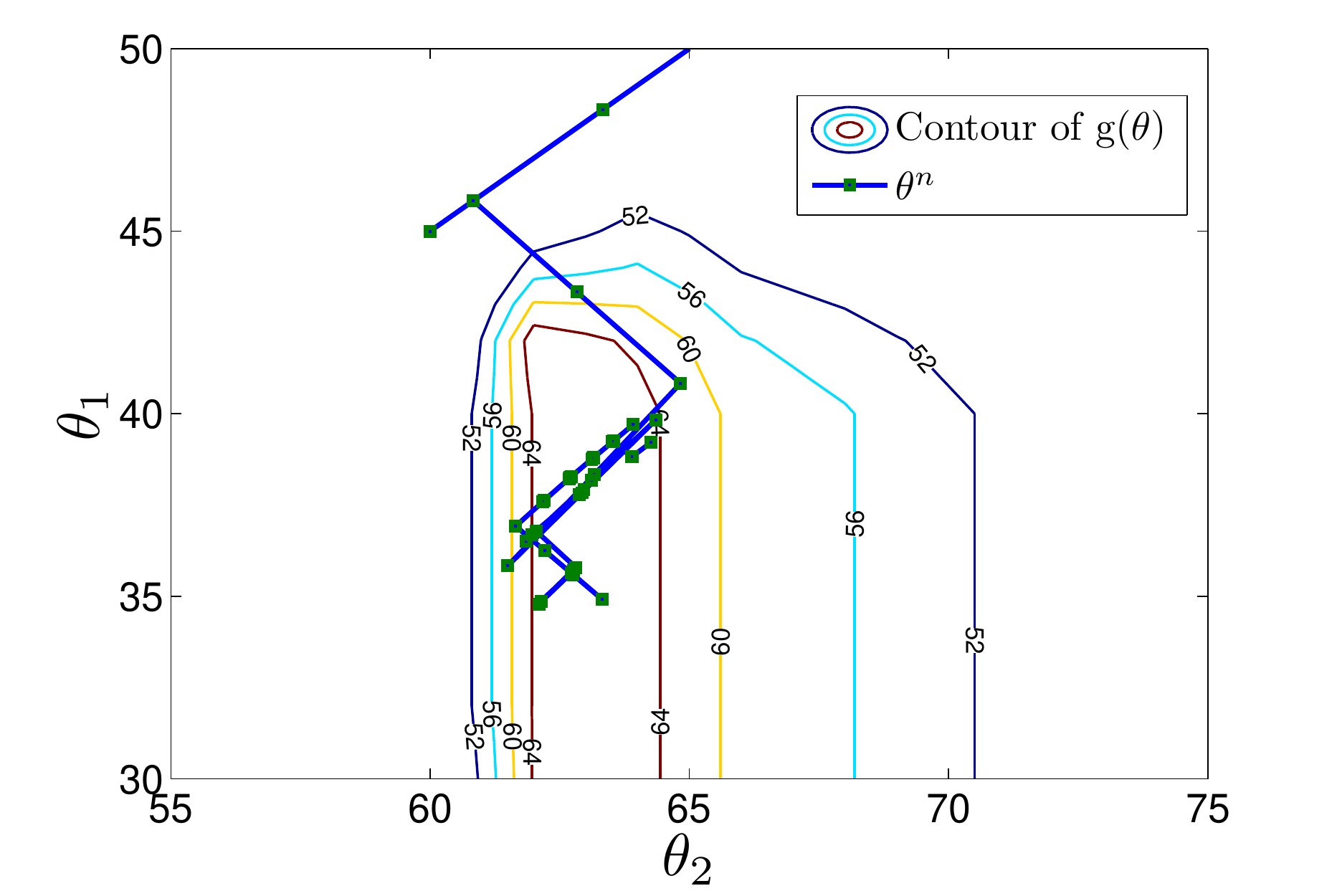}
    \label{fig:experiment4_detail}}
    \caption{The updated threshold vector $\boldsymbol{\theta}^n$s of the proposed threshold optimization algorithm are shown in \subref{fig:experiment4_large}. The dashed box is shown blown up in \subref{fig:experiment4_detail} to illustrate more detail. The contours of $\mathrm{g}(\boldsymbol{\theta})$ are also shown on the figure for reference.}
\label{fig:experiment4_fix_alpha}
\end{figure}

In Fig.~\ref{fig:experiment4_fix_alpha}, we plot the threshold vectors $\boldsymbol{\theta}^n$ in the proposed threshold optimizing algorithm when the channel scaling parameter is fixed at $\gamma=6$. We set the initial updating step-size $\epsilon_j^0=10,~\forall j$. It is apparent that the threshold vector converges quickly to the area where $\mathrm{g}(\boldsymbol{\theta})$ is maximized.

\section{Conclusions and Future Work}
\label{sec:conclusion}
We created a new QoE metric based on the second-order empirical cumulative distribution function (eCDF) of time-varying video quality. We then proposed an online rate adaptation algorithm to maximize the percentage of video users who satisfy the QoE constraints on the second-order cumulative distribution function. Furthermore, we devised a threshold-based admission control strategy that blocks new video users whose QoE constraints cannot be satisfied. Simulation results showed that combining the proposed approaches leads to a 40\% reduction in wireless network resource consumption.

The users' video quality expectation play a critical role in our QoE metric. It is important to observe that our subjective study was conducted in a controlled environment. In reality, however, users' expectations for video quality may depend on various factors in the environment (e.g., user mobility, device type, lighting conditions). In the future, we plan to conduct subjective study in more diversified, ``worldly" environments to obtain a better understanding of and an improved ability to predict users' quality expectations.

\appendices
\section{Proof of Theorem 1}
\label{append:2}
\begin{IEEEproof}
Since $\left[ x-\mathrm{q}(t)\right]^+$ is a convex function of $x$, $\mathrm{F}^{(2)}(x;\mathrm{q})$ is a linear combination of $\left[ x-\mathrm{q}(t)\right]^+$ and is thus also a convex function of $x$. Without loss of generality, assume function $\mathrm{h}(x)$ is also convex\footnote{Otherwise, we can simply replace $\mathrm{h}(x)$ with another function whose epigraph is the convex hull of $\mathrm{h}(x)$'s epigraph.}. Let $x_i<x_j$, where $i,j\in\mathcal{I}$. If \eqref{eq:unknown} is satisfied, then $\mathrm{F}^{(2)}(x_i;\mathrm{q})\leq \mathrm{h}(x_i)$ and $\mathrm{F}^{(2)}(x_j;\mathrm{q})\leq \mathrm{h}(x_j)$. For any $\lambda\in[0,1]$ and $x=\lambda x_i+(1-\lambda)x_j$, we have $\mathrm{F}^{(2)}(x;\mathrm{q})=\mathrm{F}^{(2)}(\lambda x_i+(1-\lambda)x_j;\mathrm{q})\leq\lambda \mathrm{F}^{(2)}(x_i;\mathrm{q})+(1-\lambda)\mathrm{F}^{(2)}(x_j;\mathrm{q})\leq\lambda \mathrm{h}(x_i)+(1-\lambda)\mathrm{h}(x_j)=\bar{\mathrm{h}}(x)$. Because $[0,100]$ is a compact set, the convexity of $\mathrm{h}(x)$ implies its continuity. Therefore, $\mathrm{h}(x)$ can be approximated by piece-wise linear functions to arbitrary accuracy.
\end{IEEEproof}
\section{Proof of Theorem 2}
\label{append:1}
\begin{IEEEproof}
Note that {Algorithm~\ref{alg:online_threshold_case1}} can be viewed as a stochastic approximation algorithm \cite{SA} with an associated mean ordinary differential equation (ODE)
\begin{equation}
\label{eq:ode}
\frac{\text{d}\theta(t)}{\text{d}t}=\mathbb{E}[\mathsf{Y}(\theta(t))],
\end{equation}
where $\mathsf{Y}(\theta)$ is a random variable that denotes the updating direction when the threshold is $\theta$, we have
\begin{align}
\mathbb{E}[\mathsf{Y}(\theta(t))]=1-2\mathrm{p}^L(\theta(t)).
\end{align}
According to Assumption~\ref{asump:p_k}, there exists a unique $\theta'$ such that $\mathrm{p}^L(\theta')=1/2$ and $\mathbb{E}[\mathsf{Y}(\theta')]=0$. By the monotonicity of $\mathrm{p}^L(\theta)$, we have $\mathbb{E}[\mathsf{Y}(\theta)]>0,~\forall \theta< \theta'$ and $\mathbb{E}[\mathsf{Y}(\theta)]<0,~\forall \theta> \theta'$. If we define a function $\mathrm{V}(\theta)=1/2(\theta-\theta')^2$, then
\begin{align*}
\frac{\text{d}\mathrm{V}(\theta(t))}{\text{d}t}&=(\theta(t)-\theta')\frac{\text{d}\theta(t)}{\text{d}t}\\
&=(\theta(t)-\theta')\left(1-2\mathrm{p}^L(\theta(t))\right)\\
&=-(\theta(t)-\theta')\left(2\mathrm{p}^L(\theta(t))-2\mathrm{p}^L(\theta')\right)\\
&=-2(\theta(t)-\theta')\left(\mathrm{p}^L(\theta(t))-\mathrm{p}^L(\theta')\right)
\end{align*}
For $\forall \theta<\theta'$, we have $\mathrm{p}^L(\theta)-\mathrm{p}^L(\theta')\leq M(\theta-\theta')$.  For $\forall \theta>\theta'$, $\mathrm{p}^L(\theta)-\mathrm{p}^L(\theta')\geq M(\theta-\theta')$. In sum, we have
\begin{align*}
\frac{\text{d}\mathrm{V}(\theta(t))}{\text{d}t}\leq-2M(\theta(t)-\theta')^2.
\end{align*}
By Theorem 5.4.1 in \cite[p.145]{SA}, we have $\lim_{n\rightarrow\infty}\theta^n=\theta'$. Next, we prove that $\theta'\in[\theta^*-\delta,\theta^*)$.

We define a binary random variable $\mathsf{S}_u$ such that $\mathsf{S}_u=1$ if video user $u$ satisfies the QoE constraints $\{\mathrm{F}^\mathrm{(2)}(x_i;\mathrm{q}_u)\leq h_i, \forall x_i\in \mathcal{I}\}$. Otherwise, we define $\mathsf{S}_u=0$. Denote by $\mathcal{U}^L=\{u_1,\dots,u_L\}$ the indices of the $L$ admitted video users in an iteration of Algorithm~\ref{alg:online_ac_case1}. Let $\pi_\theta$ be the joint distribution of the variables $\{\mathsf{S}_u,~\forall u\in \mathcal{U}^L\}$ when the admission threshold is $\theta$. Then, we have
\begin{equation}
\label{eq:them1_eq1}
\begin{aligned}
\mathrm{p}^L(\theta)&=\mathbb{P}^{\pi_\theta}\left(\mathsf{S}_u=1,~\forall u\in \mathcal{U}^L\right)\\
&=\mathbb{P}^{\pi_\theta}\left(\mathsf{S}_{u_1}=1\right)\mathbb{P}^{\pi_\theta}\left(\mathsf{S}_{u_\ell}=1,~2\leq \ell\leq L|\mathsf{S}_{u_1}=1\right).
\end{aligned}
\end{equation}
Since the users are competing with each other for network resources, if the QoE constraints of a video user are satisfied, the probability of satisfying other users' QoE constraints is reduced. Thus,
\begin{equation}
\label{eq:them1_eq2}
\begin{aligned}
&\mathbb{P}^{\pi_\theta}(\mathsf{S}_{u_\ell}=1,~\forall 2\leq \ell\leq L|\mathsf{S}_1=1)\\
\leq&\mathbb{P}^{\pi_\theta}(\mathsf{S}_{u_\ell}=1,~\forall 2\leq \ell\leq L).
\end{aligned}
\end{equation}
Substitute \eqref{eq:them1_eq2} into \eqref{eq:them1_eq1} yields
\begin{equation}
\label{eq:them1_eq3}
\begin{aligned}
\mathrm{p}^L(\theta)&\leq\mathbb{P}^{\pi_\theta}(\mathsf{S}_{u_1}=1)\mathbb{P}^{\pi_\theta}(\mathsf{S}_{u_\ell}=1,~\forall 2\leq \ell\leq L)\\
&\leq\mathbb{P}^{\pi_\theta}(\mathsf{S}_{u_1}=1)\mathbb{P}^{\pi_\theta}(\mathsf{S}_{u_2}=1)\mathbb{P}^{\pi_\theta}(\mathsf{S}_{u_\ell}=1,~\forall 3\leq \ell\leq L)\\
&\leq\Pi_{\ell=1}^L\mathbb{P}^{\pi_\theta}(\mathsf{S}_{u_\ell}=1)\\
&=\left(1-\mathrm{e}(\theta)\right)^L.
\end{aligned}
\end{equation}
Since $L\geq\frac{-\log2}{\log(1-\mathrm{e}(\theta^*-\delta))}$, it follows that
$\mathrm{p}^L(\theta^*-\delta)\leq\left(1-\mathrm{e}(\theta^*-\delta)\right)^L\leq1/2$. Because of the monotonicity of $\mathrm{p}^L(\theta)$, we know that $\theta^*-\delta\leq \theta'$. Moreover, because $\mathrm{p}^L(\theta')=1/2>0$, we have  $\theta'<\theta^*$.
\end{IEEEproof}

\ifCLASSOPTIONcaptionsoff
  \newpage
\fi
\bibliographystyle{IEEEtran}
\bibliography{IEEEabrv,allbib}
\end{document}